\documentclass[sn-nature,Numbered]{sn-ww}
\usepackage{graphicx}%
\usepackage{multirow}%
\usepackage{amsmath,amssymb,amsfonts}%
\usepackage{amsthm}%
\usepackage{mathrsfs}%
\usepackage[title]{appendix}%
\usepackage{xcolor}%
\usepackage{textcomp}%
\usepackage{manyfoot}%
\usepackage{booktabs}%
\usepackage{algorithm}%
\usepackage{algorithmicx}%
\usepackage{algpseudocode}%
\usepackage{listings}%

\usepackage[super]{cite}

\begin{document}

\title[A Giant Disk at z$\sim$3]{A Giant Disk Galaxy Two Billion Years After The Big Bang}

\author[1]{\sur{Weichen} \fnm{Wang}}
\email{weichen.wang@unimib.it}

\affil[1]{\orgdiv{Department of Physics}, \orgname{Universita degli Studi di Milano-Bicocca}, \orgaddress{\street{Piazza della Scienza, 3}, \city{Milano}, \postcode{I-20126}, \country{Italy}}}

\author[1]{\sur{Sebastiano} \fnm{Cantalupo}}

\author[1]{\sur{Antonio} \fnm{Pensabene}} 

\author[1]{\sur{Marta} \fnm{Galbiati}}

\author[1]{\sur{Andrea} \fnm{Travascio}}

\author[2]{\sur{Charles C.} \fnm{Steidel}}

\affil[2]{\orgdiv{Cahill Center for Astronomy and Astrophysics}, \orgname{California Institute of Technology}, \orgaddress{ \street{1200 E California Blvd, MC 249-17}, \city{Pasadena}, \postcode{CA 91125}, \country{USA}}}

\author[3]{ \sur{Michael V.} \fnm{Maseda} }
\affil[3]{\orgdiv{Department of Astronomy}, \orgname{University of Wisconsin-Madison}, \orgaddress{ \street{475 N. Charter Street}, \city{Madison}, \postcode{WI 53706}, \country{USA}}}

\author[4]{ \sur{Gabriele} \fnm{Pezzulli} }
\affil[4]{\orgdiv{Kapteyn Astronomical Institute}, \orgname{University of Groningen}, \orgaddress{ \street{Landleven 12}, \postcode{9747 AD}, \city{Groningen}, \country{The Netherlands}}}

\author[1]{\sur{Stephanie} \fnm{de Beer}}

\author[1,5]{\sur{Matteo} \fnm{Fossati}}
\affil[5]{\orgdiv{INAF - Osservatorio Astronomico di Brera}, \orgaddress{\street{via Bianchi 46}, \city{Merate (LC)}, \postcode{I-23087}, \country{Italy}}}

\author[1,6]{\sur{Michele} \fnm{Fumagalli}}
\affil[6]{\orgdiv{INAF - Osservatorio Astronomico di Trieste}, \orgaddress{\street{via G. B. Tiepolo 11}, \city{Trieste}, \postcode{I-34143}, \country{Italy}}}

\author[7]{\sur{Sofia G.} \fnm{Gallego}}
\affil[7]{\orgdiv{Astroparticle and Cosmology Laboratory}, \orgname{Université Paris Cité}, \orgaddress{\city{Paris}, \postcode{F-75013}, \country{France}}}

\author[1]{\sur{Titouan} \fnm{Lazeyras}}

\author[8]{\sur{Ruari} \fnm{Mackenzie}}
\affil[8]{\orgdiv{Department of Physics}, \orgname{ETH Zürich}, \orgaddress{ \street{Wolfgang-Pauli-Strasse 27}, \postcode{8093} \city{Zürich}, \country{Switzerland}}}

\author[9]{ \sur{Jorryt} \fnm{Matthee}}
\affil[9]{\orgname{Institute of Science and Technology Austria (ISTA)}, \orgaddress{ \street{Am Campus 1}, \postcode{3400} \city{Klosterneuburg}, \country{Austria}}}

\author[10]{\sur{Themiya} \fnm{Nanayakkara}} 
\affil[10]{\orgdiv{Centre for Astrophysics and Supercomputing}, \orgname{Swinburne University of Technology}, \orgaddress{ \street{P.O. Box 218}, \city{Hawthorn}, \postcode{VIC 3122}, \country{Australia}}}

\author[1]{\sur{Giada} \fnm{Quadri}}

\abstract{\bf 

Observational studies showed that galaxy disks are already in place in the first few billion years of the universe \cite{Genzel2006,Law2012,Genzel2017,Rizzo2020}. 
The early disks detected so far, with typical half-light radii of 3 kiloparsecs at stellar masses around $\tt 10^{11} M_\odot$ for redshift $z\tt\sim3$, are significantly smaller than today's disks with similar masses\cite{vanderwel2014b,Ogle2016,ForsterSchreiber2020,Ward2023,Varadaraj2024}, in agreement with expectations from current galaxy  models\cite{MMW1998,Costantin2023,LaChance2024}.
Here, we report observations of a giant disk at $z\tt =3.25$, when the universe was only 2 billion years old, with a half-light radius of 9.6 kiloparsecs and stellar mass of $\tt  3.7^{+2.6}_{-2.2}\times 10^{11} M_\odot$. 
This galaxy is larger than any other kinematically-confirmed disks at similar epochs and surprisingly similar to today's largest disks regarding size and mass. JWST imaging and spectroscopy reveal its spiral morphology and a rotational velocity consistent with local Tully-Fisher relation\cite{Lelli2016}.
Multi-wavelength observations show that it lies in an exceptionally dense environment, where the galaxy number density is over ten times higher than the cosmic average and mergers are frequent. 
The discovery of such a giant disk suggests the presence of favorable physical conditions for large-disk formation in dense environments in the early universe, 
which may include efficient accretion of gas carrying coherent angular momentum and non-destructive mergers between exceptionally gas-rich progenitor galaxies. 
}

\maketitle

\begin{figure}[h]
\centering
\includegraphics[width = 5.15in]{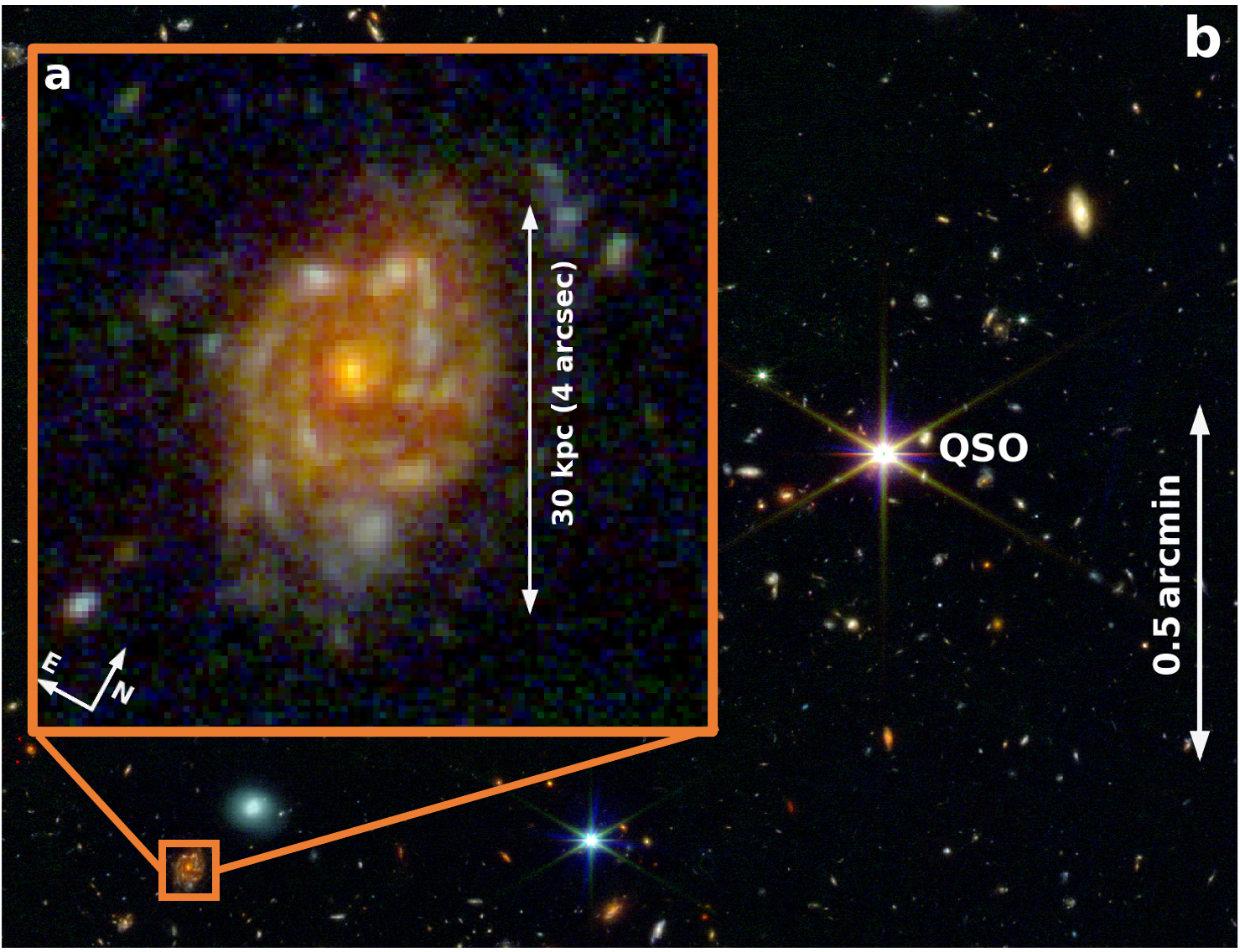}
\caption{ {\bf Composite false-color images of the Big Wheel galaxy at $z=3.245$ (a) and its surrounding area (b).} The galaxy shows a red and compact center and a giant stellar disk extending to at least 30\,kiloparsecs in diameter. The disk appears clumpy with manifest spiral structures {\bf (a)}. The Big Wheel is located about 70 arcsec away (about 0.5 pMpc) from a bright quasar at a similar redshift {\bf (b)}. The quasar was originally chosen as the center of the observation field. This region shows an exceptionally high galaxy number density compared to the cosmic average. The filters used to create the color image are HST F814W (0.8$\,\mu$m; blue), JWST F150W2 (1.5$\,\mu m$; green), and JWST F322W2 (3.2$\,\mu$m; red). The disk is visible in the green and red channels but not in the blue channel. }\label{fig:fig1}

\end{figure}

The galaxy, dubbed ``Big Wheel", was discovered serendipitously in a bright-quasar field at redshift ($z$) of 3.25 \cite{Pensabene2024} 
 through James Webb Space Telescope (JWST) imaging at wavelengths of 1.5$\,\mu$m and 3.2$\,\mu$m (rest-frame 0.4$\,\mu$m and 0.8$\,\mu$m).  
 Hubble Space Telescope (HST) observations at 0.8$\,\mu$m (rest-frame 0.2$\,\mu$m) reveal only isolated clumps at the galaxy outskirts, possibly tracing young stars and/or low dust obscuration. A RGB-image combining the HST and JWST images is shown in  Fig.~\ref{fig:fig1}b,
 while Fig.~\ref{fig:fig1}a shows a zoomed-in view of the Big Wheel.  
 The Big Wheel features a red center, visible only in the JWST near-infrared filters (green and red channels), and a stellar disk extending to at least 30\,kiloparsecs in diameter. 
 Spiral-arm features are visible, appearing clumpy in rest-frame ultraviolet, reminiscent of some spiral galaxies at $z \sim 0$ \cite{GildePaz2007}.  Its stellar half-light radius is 9.6\,kiloparsecs along the major axis in the rest-frame optical (0.5$\,\mu m$), as measured using {\sc statmorph} \cite{Rodriguez-Gomez2019} (see \emph{Methods}).

\begin{figure}
\centering
\includegraphics[width = 5.1 in]{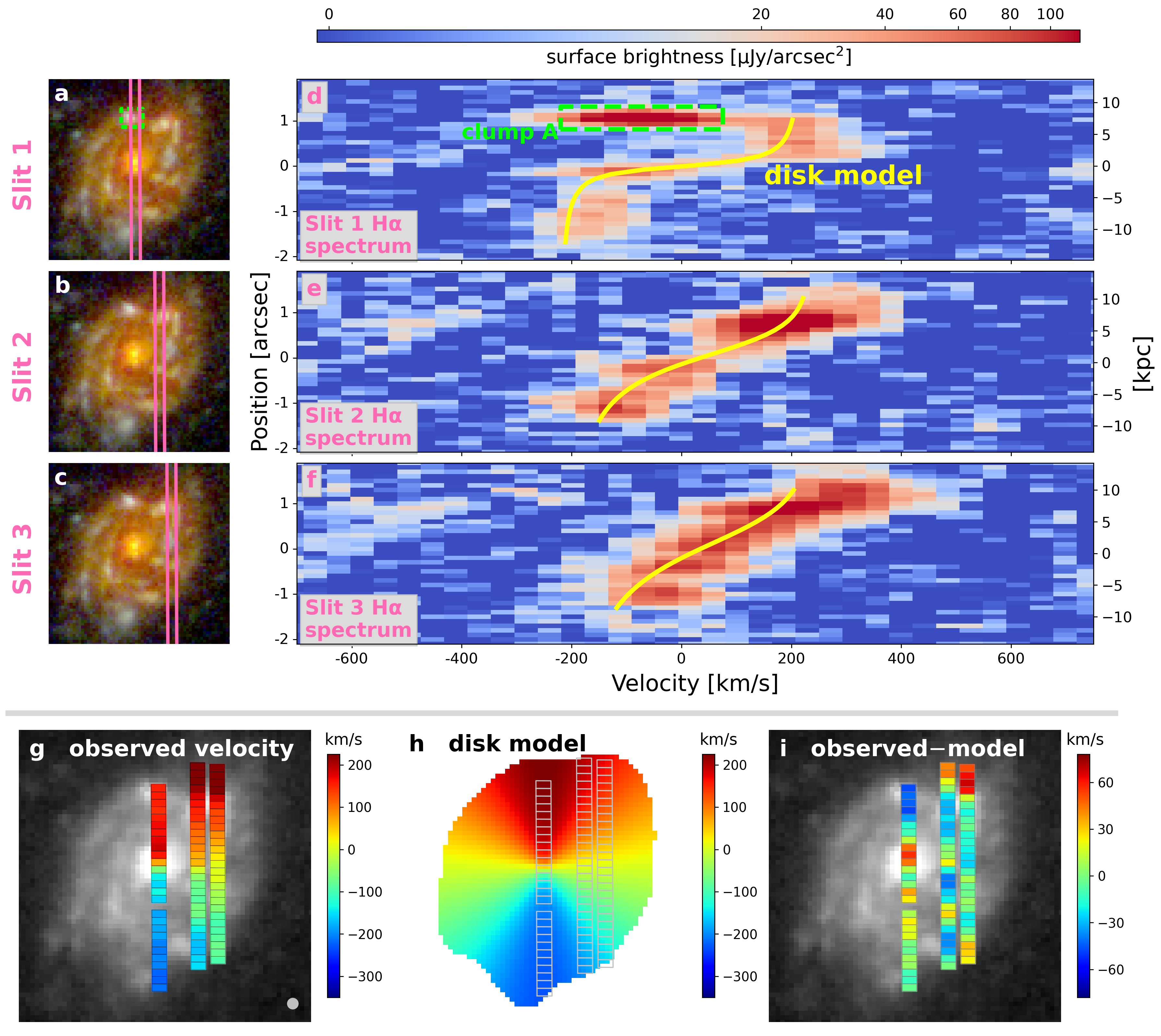}
\caption{ {\bf JWST kinematic measurements supporting the rotating-disk nature of the Big Wheel.} The three spectral slits used in the observations are shown in {\bf a-c}. These panels are aligned with the 2D spectra in {\bf d-f} along the Y direction. The 2D spectra are continuum-subtracted and centered on the H$\alpha$ emission line, where the X axis is the dispersion direction indicating the line-of-sight velocity and the Y axis indicates the spatial position along the slit. The yellow curve overlaid on each spectrum represents the ridge-line of the H$\alpha$ profile predicted by a simple disk model. The velocity map in {\bf g} is obtained by measuring the centroids of the H$\alpha$ profiles across the three slits, excluding a clump with peculiar morphology and kinematics (green box) and a region without H$\alpha$ detection in Slit 1. The map from observations is in agreement with the disk model {\bf (h)} with no significant residuals  over the main stellar disk body {\bf (i)}. Small residuals ($\simeq$ 30-60 km/s) in the kinematics are only present at the upper ends of Slits 2 and 3, at the edge or outside of the stellar disk {\bf (i)}.  The spatial resolution of the observations is indicated at the bottom right of {\bf g}. }\label{fig:fig2}
\end{figure}

Hydrogen H$\alpha$ spectroscopy obtained with the Near Infrared Spectrograph (NIRSpec) confirms the rotating-disk nature of Big Wheel.
Three spectroscopic slits (see \emph{Methods}) were placed onto the galaxy (Fig.~\ref{fig:fig2}a-c), including one covering the galaxy center. %
H$\alpha$ two-dimensional spectra are presented in Fig.~\ref{fig:fig2}d-f, whereas the corresponding slit positions are indicated in Fig.~\ref{fig:fig2}a-c. 
Initial evidence supporting disk rotation can be gleaned from the slit spectrum covering the galaxy center (Fig.~\ref{fig:fig2}d).
Moving towards to the external regions, the velocity increases first and then plateaus at around $\pm 200$\,km/s (prior to inclination correction). Such pattern is typical of disk galaxies with flat rotation curves \cite{Lelli2016}. 
Additional features are present in the spectra, such as a bright clump which appears kinematically distinct from the disk (green box) and could be associated with a companion galaxy seen in projection.

The H$\alpha$ spectra of the remaining slits (Fig.~\ref{fig:fig2}e-f) also have patterns consistent with disk-like rotation. To demonstrate this, we measure the rotation velocities from the H$\alpha$ profiles across the slits as shown in Fig.~\ref{fig:fig2}d-f (see \emph{Methods}) and present the result as a velocity map in Fig.~\ref{fig:fig2}g. 
As shown in Fig.~\ref{fig:fig2}d-f, these profiles
agree remarkably well over the main stellar disk body
with predictions from a simple disk kinematic model, in which a flat rotation curve is adopted (yellow curves in Fig.~\ref{fig:fig2}d-f).

The disk model, described in detail in \emph{Methods}, includes six free parameters which specify the disk center location, major axis orientation, inclination, and normalization of the rotation curve. The model was fit to the data using the Markov Chain Monte Carlo scheme {\sc{emcee}} \cite{Foreman-Mackey2013}. The best fit model is presented in Fig.~\ref{fig:fig2}h and the residuals in Fig.~\ref{fig:fig2}i.  The maximum rotation velocity (corrected for inclination) $v_\mathrm{rot}$ is $280^{+32}_{-31}$\,km/s, i.e. $4.5^{+1.4}_{-1.1}$ times the velocity dispersion $\sigma_\mathrm{int}$ ($61^{+17}_{-13}$ km/s; see \emph{Methods}). These values indicate that the Big Wheel is rotationally supported and has a dispersion value consistent with (smaller) turbulent disks at similar and lower redshifts \cite{Kassin2012,ForsterSchreiber2020}. Combining $v_\mathrm{rot}$ and $\sigma_\mathrm{int}$ using the commonly adopted relation in the literature\cite{Ubler2017} (see \emph{Methods}) we obtain a circular velocity $v_\mathrm{circ}=304^{+32}_{-31}\,$km/s.
The velocity map of the molecular gas obtained through ALMA observations \cite{Pensabene2024} covering the full galaxy, albeit with lower spatial resolution, shows consistent results (see \emph{Methods}).

\begin{table}[h]
\caption{\bf Properties of the Big Wheel galaxy}\label{tab:table1}%
\centering
\begin{tabular}{@{}lll@{}}
\toprule
Property & Value & Unit \\
\midrule
Right ascension (J2000) & 00h41m35.113s & - \\
Declination (J2000) & -49d37m12.42s & - \\
Redshift\footnotemark[1] & 3.2452 & - \\
Stellar mass\footnotemark[2] & $3.7^{+2.6}_{-2.2}\times 10^{11}$ & solar mass ($M_\odot$) \\
Stellar mass (assuming a parametric SFH)\footnotemark[3] & $1.7^{+1.4}_{-0.7}\times 10^{11}$ & solar mass ($M_\odot$) \\
Star-formation rate\footnotemark[2] & $2.5^{+7.5}_{-2.1}\times 10^{2}$ & $M_\odot/$yr \\
Half-light radius\footnotemark[4] & $1.27^{+0.07}_{-0.16}$, $9.6^{+0.5}_{-1.2}$ & arcsec, kiloparsec \\
Minor-to-major axis ratio & $0.69^{+0.03}_{-0.01}$ & - \\
Disk inclination angle\footnotemark[5] & $48^{+1}_{-4}$ \  (morphology) & degree \\
 & 52$_{-5}^{+5}\,\,$ (kinematics) & degree \\
H$\alpha$ rotational velocity\footnotemark[6] ($v_\mathrm{rot}$) & $280^{+32}_{-31}$ & km/s \\
H$\alpha$ velocity dispersion ($\sigma_\mathrm{int}$) & 61$^{+17}_{-13}$ & km/s \\
$v_\mathrm{rot}/\sigma_\mathrm{int}$ & 4.5$^{+1.4}_{-1.1}$ & - \\
Circular velocity\footnotemark[6] ($v_\mathrm{circ}$) & $304^{+32}_{-31}$ & km/s \\
CO(4-3) line luminosity & $1.3\times 10^8$ & solar luminosity ($L_\odot$) \\
H$_2$ mass estimated from CO & $1.8^{+1.0}_{-0.8}\times 10^{11}$ & $M_\odot$ \\
Environment overdensity\footnotemark[7] & $\geq 10$ & - \\
\botrule
\end{tabular}
\footnotetext[1]{Measured from the H$\alpha$ and [N II] emission lines.} 
\footnotetext[2]{Values inferred from SED fitting with a non-parametric star formation history using {\sc prospector} \cite{Leja2017}. and adopted in this work. The star-formation rate is calculated from a 100\,Myr timescale.} 
\footnotetext[3]{  Stellar mass inferred using an alternative parametric form of star formation history (SFH) for reference. See \emph{Methods} for details.}
\footnotetext[4]{Measured along the major axis of the galaxy at rest-frame 0.5$\,\mu$m.}
\footnotetext[5]{Angle between the galaxy disk plane and sky plane, with 0 degree indicating fully face-on and 90 degrees fully edge-on.}
\footnotetext[6]{See \emph{Methods} for details regarding the definition of the two quantities.}
\footnotetext[7]{Inferred from galaxy number density measurements as described in [\citen{Pensabene2024}].}
\end{table}


The stellar mass and star formation rate (SFR) were obtained by fitting the photometry in seven filters against spectral energy distribution (SED) models with the {\sc prospector} tool \cite{Leja2017} (see \emph{Methods}). 
Flexible star formation histories (SFHs) and dust attenuation laws were adopted, as well as the potential emission from an active galactic nucleus (AGN). The best-fit stellar mass and SFR values are $3.7^{+2.6}_{-2.2}\times10^{11}\,M_\odot$ and $2.5^{+7.5}_{-2.1} \times 10^2 \,M_\odot/$yr, respectively. These results indicate that the Big Wheel is one of the most massive systems detected so far at $z\gtrsim 3$, some of which discovered only recently with JWST \cite{Nanayakkara2024}.

\begin{figure}[t]
\centering
\includegraphics[width = 5.15in]{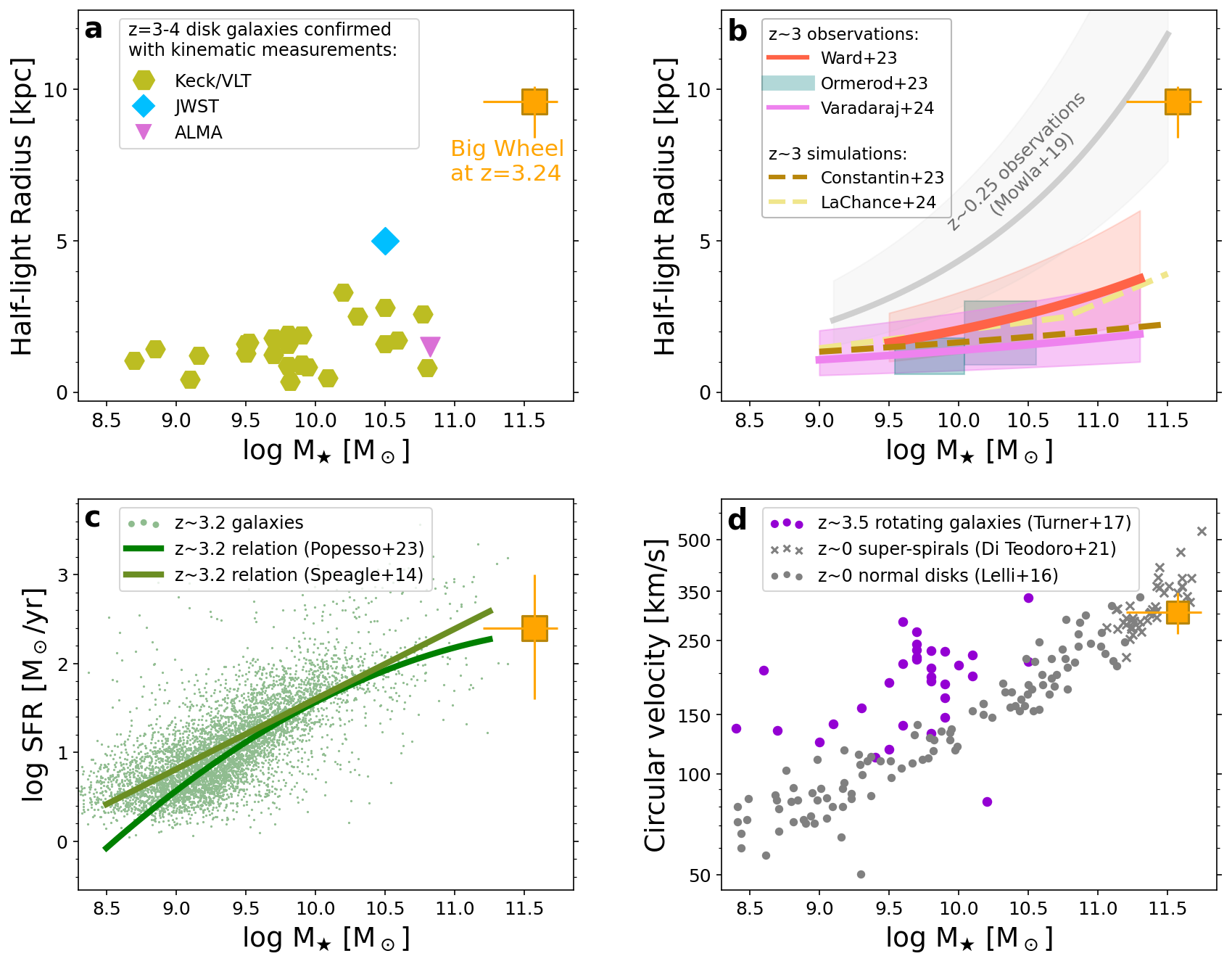}
\caption{ {\bf Physical properties of the Big Wheel compared to other galaxies at similar and lower redshifts.} The Big Wheel galaxy, at $z= 3.245$, is larger and more massive than any other kinematically-confirmed disk galaxy discovered to date at similar redshifts {\bf (a)}. It is well above the extrapolation of the size-mass relation at a mass of $10^{11.5}$ M$_\odot$ and $z\sim 3$ from observations and simulations and more consistent with the size of $z \sim 0$ galaxies {\bf (b)}. Furthermore, it follows the Tully-Fisher relation (circular velocity versus stellar mass) of the disk galaxies at $z\sim 0$ {\bf (d)}. The Big Wheel has an SFR consistent with the (extrapolated) trend with stellar mass for typical star-forming galaxies at $z \sim 3$ {\bf (c)}. Errorbars of the stellar mass, radius, SFR, and circular velocity take into account both random and systematic uncertainties, which are described in \emph{Methods}. Details of the literature studies included here are also provided in \emph{Methods}. \label{fig:fig4}}
\end{figure}

The physical properties of the Big Wheel are summarized in Tab.~\ref{tab:table1} and are compared to the observed galaxy population at similar redshifts in Fig.~\ref{fig:fig4}.  
Most intriguingly, the Big Wheel has a larger radius (measured at rest-frame 0.5$\,\mu m$) and is more massive than any other kinematically-confirmed disk galaxy discovered so far at similar redshifts (see Fig.~\ref{fig:fig4}a and \emph{Methods}). 
Fig.~\ref{fig:fig4}b shows the size versus stellar mass relations for $z \sim 3$ star-forming galaxies for both observations and simulations, as well as the observed relation at $z\sim 0$. The Big Wheel is clearly offset to high values, both in terms of stellar mass and sizes, with respect to both the observed and simulated relations at $z \sim 3$. 
In addition, in terms of size, the Big Wheel galaxy is more similar to $z \sim 0$ disk galaxies than other currently known disk galaxies at $z \sim 3$. As shown in Fig.~\ref{fig:fig4}d, it is located on top of the stellar mass Tully-Fisher relation of the $z\sim 0$ disks \cite{Lelli2016} 
with a stellar mass and circular velocity similar to local ``super-spirals''\cite{DiTeodoro2021}. 
Despite the similarities to the $z\sim 0$ objects, the Big Wheel galaxy is actively growing in mass like other galaxies at $z\sim3$, with an SFR consistent with the trend of typical $z \sim 3$ star-forming galaxies (Fig.~\ref{fig:fig4}c).

Our observations reveal that the Big Wheel is a giant rotating disk with physical properties unique for $z\sim 3$, raising questions about its formation scenario. 
Within the classical theoretical framework \cite{MMW1998,Bullock2001}, the disk size is 
expected to be simply proportional to the halo size times the dimensionless halo spin parameter ($\lambda$)\cite{MMW1998,Bullock2001}, with small deviations with halo concentration and disk-to-halo mass ratio.  The halo $\lambda$ follows a distribution well constrained from simulations and can be fit with a log-normal function with an average of $\simeq0.035$ and a log-normal standard deviation (in log base 10) of $\simeq0.25$\cite{Bullock2001,Bryan2013} without significant variations with halo mass and redshift.\cite{Bryan2013,Zjupa2017}. 

Despite the model simplicity, its predictions are consistent with the observed galaxy size distributions
both in terms of shape and scatter.
In particular, the most recent observations at $z\simeq3$, find a log-normal size distribution with an intrinsic scatter of $\simeq0.22-0.28$\cite{Ward2023,Varadaraj2024}, which is strikingly similar to the simple expectations mentioned.
As shown in Fig.~\ref{fig:fig4}b, the Big Wheel is at least three times larger than the expected size of star-forming, disk galaxies at its mass and redshift given the observed size-mass relation in random fields\cite{Ward2023,Varadaraj2024}.
The probability of randomly finding such a galaxy, if environment does not play a role, 
is less than 2\% (see \emph{Methods} for details). The Big Wheel is one of the few star forming galaxies with rest-frame optical size measurement and a mass above 10$^{11}$ $M_\odot$ found so far at $z>3$, and the only one with confirmed kinematical measurements demonstrating its rotating disk nature. As such, its serendipitous discovery in one of the largest overdensities of galaxies found so far at $z\gtrsim3$ \cite{Pensabene2024} suggests that additional physical mechanisms could be at play in determining the size of massive disk galaxies in such regions of the universe.

Major mergers are expected to be more frequent than the cosmic average in overdense regions\cite{Jian2012} and
a few models suggest that they can, in exceptional conditions, facilitate disk growth by increasing the disk spin rather than destroying it.\cite{Governato2009,Hopkins2009}
 In particular, models suggest that disks can survive disruption or reform afterwards if mergers have favorable orbital parameters and
 the progenitor galaxies are gas-rich\cite{Governato2009,Hopkins2009}. 
 If these predictions are correct, the presence of such a giant disk in a large overdensity of galaxies could imply, e.g., a connection between the dense environment and an elevation of the gas fraction of galaxies. In turn, the high gas content of galaxies could be caused by more efficient accretion of the gas from the Cosmic Web in denser environments at early cosmic epochs. 
 Alternatively, large disks could also be the result of accretion of cosmic gas with coherent angular momentum, resulting in larger disk-to-halo angular momentum ratio with respect to expectations from previous analytical models \cite{Stewart2017}. 
 The relevant galaxy formation and evolution mechanisms are still not well constrained to date. To the best of our knowledge, current cosmological simulations \cite{Costantin2023,LaChance2024}, some of which are presented in Fig. \ref{fig:fig4}, have not predicted disks as large as the Big Wheel galaxy at $z\gtrsim3$ at comparable masses.

 In addition to its uncertain origin, the subsequent evolution of the Big Wheel also remains unknown.
 The fact that the galaxy is not growing in isolation and the presence of at least one companion galaxy as shown in Fig.~\ref{fig:fig2} could suggest future mergers responsible for an evolution in the Big Wheel properties.
 Moreover, its dense environment suggests the presence of a proto-cluster \cite{Pensabene2024}, hinting that its descendant might be one of the most massive members of today's galaxy clusters. Nevertheless, further studies are needed to understand how common giant disks like the Big Wheel are in dense environments at early cosmic epochs and whether their physical properties and number densities are consistent with the putative progenitors of today's most massive cluster galaxies.

\clearpage

\section*{Methods}\label{sec:methods}

\bmhead{Imaging observations and data reduction}

Imaging data was obtained from HST Program GO 17065 (PI: Cantalupo) and JWST Program GO 1835 (PI: Cantalupo) using the NIRCam filters, F150W2 and F322W2, each for 1632s of exposure and  HST filters ACS/WFC F625W and ACS/WFC F814W, for 10 and 12 orbits, respectively. The JWST data are reduced and combined using the official {\sc jwst} pipeline (v1.9.3; [\citen{Bushouse2022}]) with  calibration reference file version jwst\_1039.pmap. We implemented customized steps in the reduction to remove the correlated readout noises and stray light, following the practices adopted among the literature \cite{Bagley2023}. The image resolutions in terms of the full width of the half maximum (FWHM) are 0.12\,arcsec, 0.12\,arcsec, 0.05\,arcsec, and 0.11\,arcsec for F625W, F814W, F155W2, and F322W2, respectively.

We also obtained near-infrared data with the Very Large Telescope (VLT) in 2022 as part of a program in P110 (PI: Cantalupo). The High Acuity Wide field K-band Imager (HAWKI) was used in the Ground Layer Adaptive Optics mode, leading to spatial resolutions of 0.4\,arcsec in FWHM, using the filters CH4, H, and Ks. On-source exposure times were 60 min, 100 min, and 100 min, respectively. The images of each filter were reduced and combined using the official pipeline {\sc esorex} \cite{esorex2015}.

Photometry was performed on the images described above using the Source Extractor tool \cite{Bertin1996}. For the measurement, the image mosaics were resampled to a common grid of 0.06$\times$0.06 arcsec pixels using the {\sc drizzle} package \cite{Fruchter2002}. Source Extractor was run in dual mode for the space-based and ground-based data, using the F322W2 and H mosaics as detection images, respectively.

\begin{figure}[h]
\centering
\includegraphics[width = 5.15in]{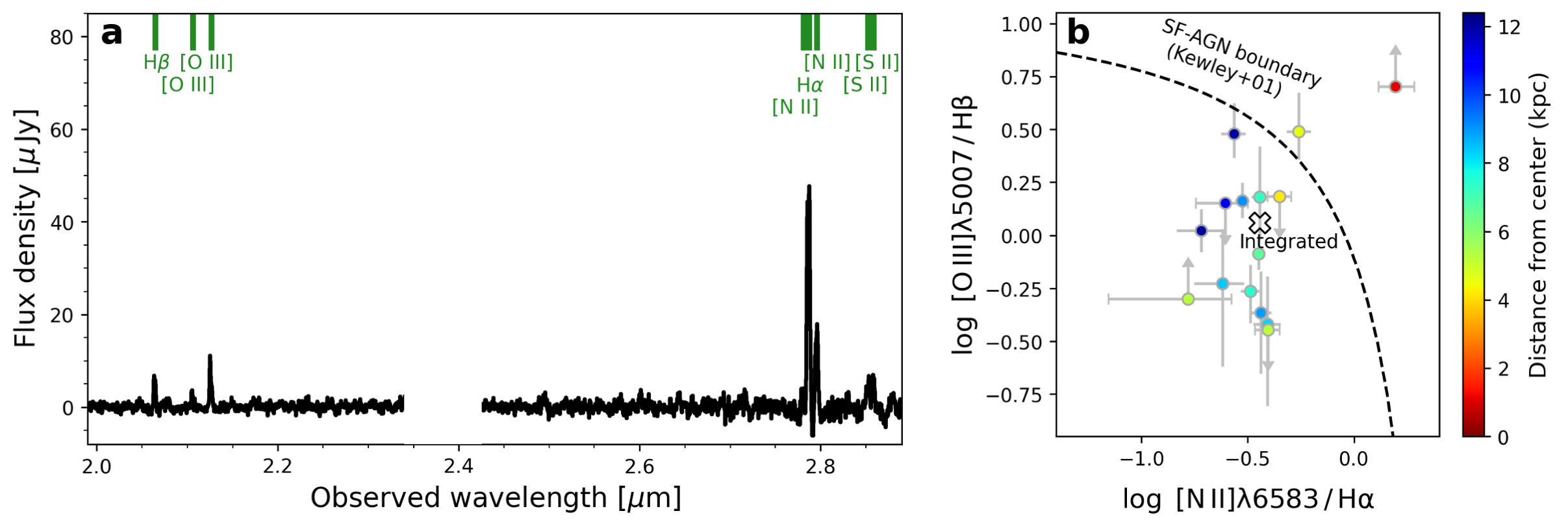}
\caption{ {\bf The integrated 1D spectrum of the Big Wheel galaxy with the continuum subtracted (a) and the spatially resolved BPT diagram (b)}. {\bf (a)} The galaxy is detected with multiple emission lines in the rest-frame optical, the observed wavelengths of which at the galaxy redshift ($z=3.245$) are indicated in the figure. The break near the middle of the figure is caused by the gap between the JWST NIRSpec detectors. {\bf (b)} The emission line ratios measured from individual parts of the galaxy within the NIRSpec slits are presented along with the integrated measurement, each color-coded by the distance from the galaxy center. Errorbars indicate 3-sigma uncertainties or detection limits. The dashed curve separates AGNs (toward upper right) from star-forming galaxies (toward lower right) and is adopted from Kewley et al \cite{Kewley2001}. Only the central part of the galaxy (r$<$1.5 kpc) appears to be dominated by AGN-like emission.  \label{fig:efig1}}
\end{figure} 

\bmhead{Spectroscopic observations and data reduction}

Follow-up spectroscopic observations using NIRSpec were obtained as a part of the same JWST program as above.
The observations used the Micro-Shutter Assembly (MSA), with the F170LP/G235H filter and grating pair, leading to a resolving power $R \sim 3200$ or equivalently 95\,km/s in  FWHM at the observed wavelength of the H$\alpha$ line ($2.8\,\mu$m). The exposure time was $\sim 3$ hours for each of the three MSA slits. The data for each slit were reduced and combined using the official {\sc jwst} pipeline (v1.11.3) with calibration reference file version jwst\_1097.pmap. The spatial resolution of the data, which is inferred with the {\sc webbpsf} tool \cite{Perrin2014}, is 0.13\,arcsec in FWHM at $2.8\,\mu$m and resampled to a pixel scale of 0.10\,arcsec.  
The smooth background and galaxy continua have been removed by
applying a median filter with a width of 3000\,km/s to each pixel row of the 2D spectra and subtracting the filtered product from the original spectra. We also extracted an integrated 1D spectrum from the combined 2D spectra. 
The resulting 1D spectrum and identified emission lines are shown in Extended Data Fig.~\ref{fig:efig1}a. The galaxy redshift is inferred from the observed wavelengths of the H$\alpha$, [N II], and H$\beta$ lines on the spectrum.

The Big Wheel was also observed with the Atacama Large {\mbox {(sub-)Millimeter}} Array (ALMA) in Cycle 8 (ID: 2021.1.00793.S; PI: Cantalupo), the details of which are provided in [14]. Data cleaning was performed down to $1.5\sigma$ with a circular mask of $2''$ in radius, which encloses emission of the galaxy in the cube. The synthesized beam size (FWHM) of the ``cleaned" data cube is $1\rlap{.}''4\times1\rlap{.}''3$ at $109.00\,{\rm GHz}$. We fit the spectral continua with zeroth-order polynomials and subtracted them from the cube using the task \texttt{imcontsub} of Common Astronomy Software Application \cite{CASA2022}.

The Big Wheel was also observed using the Advanced CCD Imaging Spectrometer onboard the Chandra X-ray Observatory during 2022-2023 (PI: Cantalupo), with a total of 634 ks of exposure. The galaxy is associated with a 3-$\sigma$ X-ray point source with a luminosity of $L_{2-10\, keV} = 9.8^{+1.8}_{-1.7} \times 10^{43}\,\mathrm{erg\cdot s}^{-1}$, which is measured assuming a photon index of 1.7-1.8 and corrected for absorption. Such a luminosity is consistent with the typical value of a moderate-luminosity Seyfert AGN \cite{Ueda2014}. The existence of the AGN was also confirmed from the ratios of the nebular emission lines from the galaxy center ($r<1.5$ kpc) on the NIRSpec spectrum, according to the  criteria by [\citen{Kewley2001,Baldwin1981}], whereas the emission lines from all other parts of the galaxy are consistent with being driven by star formation (Extended Data Fig.~\ref{fig:efig1}b).

\begin{figure}[h]
\centering
\includegraphics[width = 5.1in]{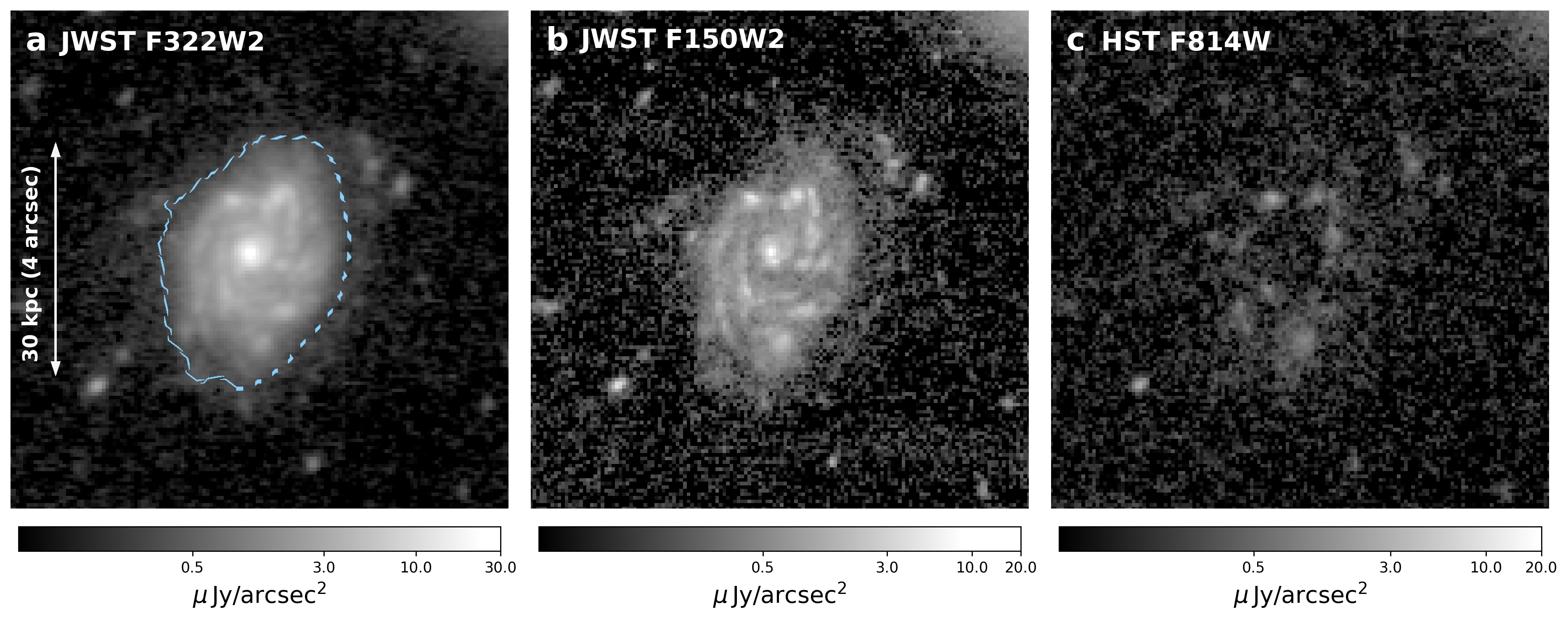}
\caption{ {\bf Galaxy broad-band images in the JWST filters F322W2 (a), F150W2 (b), and HST filter F814W (c).} Colorbars at the bottom indicate the observed surface brightness. The contour in {\bf a} represents the 5-$\sigma$ isophote demarcating the full detected extent of the galaxy, which reaches more than 30\,kiloparsecs. The galaxy disk is detected in the two near-infrared filters {\bf (a, b)}, while only a few clumps are detected in the HST filter {\bf (c)}. \label{fig:efig2}}
\end{figure} 

\bmhead{Galaxy size measurements}

The half-light radius in the rest-frame optical, is measured from the background-subtracted images in the NIRCam F150W2 and F322W2 filters. The galaxy images in these filters, as well as the F814W image, are shown in Extended Data Fig.~\ref{fig:efig2}. 
We measured the radius by running a non-parametric morphological analysis using {\sc{statmorph}}$^{16}$. This method is free from any assumption regarding the functional form of the galaxy light profile. The radius measurements were made for both the F150W2 and F322W2 filters. For each filter, we measured the radius of an ellipse enclosing half of the total light of the galaxy. 
Throughout the measurement, the radius is measured along the major axis of the ellipse and the orientation and ellipticity of the ellipse are fixed to values calculated from the mathematical moments of the pixel flux density distribution, which in turn are physically determined by the global morphology of the galaxy (see [16, \citen{Bertin1996}] for details). The half-light radii measured from the two filters are 1.34\,arcsec and 1.12\,arcsec, respectively. As a check for consistency, we also measured the radius by fitting the Sérsic\cite{Sersic1963} profiles with {\sc statmorph} using the PSFs measured from stars in the field. The resulting half-light radii are all within 5\% of the values reported above. Masking out clump A in Fig.~2a, which might be a companion galaxy, leads to only negligible ($<$2\%) changes in measured sizes.

\begin{figure}
\centering
\includegraphics[width = 5.15 in]{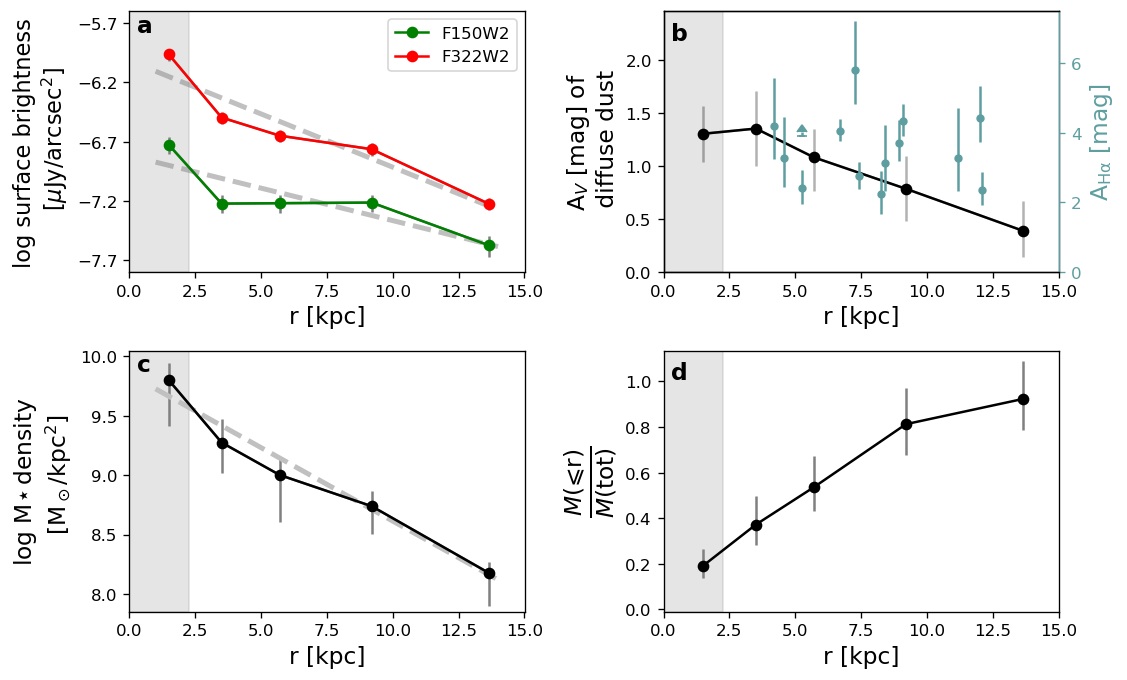}
\caption{ {\bf  Surface brightness profiles (a), along with the dust attenuation (b) and stellar mass (c, d) profiles.} 
The x-axis in each panel represents the distance from galaxy center along the major axis. {\bf (a)} Surface brightness profiles measured in two JWST filters. Dashed lines represent the profiles of perfectly exponential disks. {\bf (b)} The attenuation to stellar light due to diffuse dust (left axis, black symbols) and the H$\alpha$ line tracing young stars (right axis, blue symbols).
{\bf (c)} The stellar mass density profiles corrected for dust attenuation. The mass profile is consistent with a perfectly exponential disk profile (dashed line) everywhere within uncertainties. The half-mass radius, as can be derived from panel {\bf (d)}, is estimated to be 5.4 kpc, around ten times larger than the typical value at the same mass in $z\sim3$ simulations$^11$ but, once again, instead consistent with values measured for local galaxies\cite{Lange2015}. Shadow in each panel indicates the innermost radial bin used for SED fitting, where the AGN might contribute to the observed emission. As shown in panel d, excluding this innermost region would result in a decrease in the total mass by $<$20\%, significantly smaller than the mass uncertainty reported in Tab.~1, and also correspondingly increase the estimated half-mass radius. Errorbars represent 3-$\sigma$ uncertainties for surface brightness and A$_{\rm H\alpha}$ and 1-$\sigma$ elsewhere.
\label{fig:efig2b} }
\end{figure}

The radial light profiles, measured along the galaxy major axis, are shown in Extended Data Fig.~\ref{fig:efig2b}a together with exponential disk profiles for comparison. Both the F150W2 and F322W2 profiles are consistent with exponential disks with deviations in the central regions, which are at least partially due to the diffuse dust attenuation (panel b, left) as corroborated by the smoother dust-corrected mass density profile shown in panel c. The H$\alpha$ attenuation inferred from the H$\alpha$-to-H$\beta$ ratios following [\citen{Dominguez2013}] is also shown for reference, but we caution that it may be subject to NIRSpec flux calibration errors\cite{DEugenio2024} and traces only sightlines toward the youngest stars which does not reflect attenuation to the full continuum light seen in the two broad filters.\cite{Calzetti2001,Nelson2019}
We evaluated the uncertainties of the size measurements, which could be caused by 
beam-smearing effects due to the finite PSF of the imaging data and the imaging noise.
To quantify the effects, we created a model exponential disk with an intrinsic half-light radius of 1\,arcsec, convolved the model image with the NIRCam PSF, and added artificial noise to the image. 
Then, we measured the half-light radius from this synthetic image in the same way as was done with the real data. For each filter, such a series of steps was repeated 500 times, from which a distribution of the measured radii was obtained. We found from the distribution that the measured sizes deviate from the intrinsic value (1\,arcsec) by only 2.8\% and 3.4\% for F150W2 and F322W2 at the 3-sigma confidence level, indicating that the uncertainties due to beam-smearing and noises are relatively small. These percentage values are adopted as the relative uncertainties of our size measurements. 

In addition, we investigated whether the Big Wheel is subject to any gravitational lensing magnification caused by any foreground (i.e., low-redshift) galaxy nearby or galaxy cluster. This possibility is rejected based on the non-detection of any foreground cluster according to the X-ray and galaxy redshift measurements available to the field and the non-detection of any galaxy nearby with a mass large enough for lensing.

Finally, following the convention adopted in the literature,$^{5,8,}$\cite{Mowla2019} we inferred the half-light radius of the Big Wheel galaxy at rest-frame 0.5$\,\mu$m. It was calculated by interpolating between the radius values measured in the two NIRCam filters, under the assumption that the radius is a linear function of wavelength. To account for the systematic uncertainty due to the  interpolation conservatively, we adopted the sizes measured in the two individual filters as the bounds of the uncertainty range, and then combined in quadrature this uncertainty with the propagated uncertainty due to the beam-smearing and noise. The final calculated half-light radius is $1.27^{+0.07}_{-0.16}$\,arcsec, or equivalently $9.6^{+0.5}_{-1.2}$\,kiloparsecs.

\bmhead{Kinematic analysis and modeling}

\begin{figure}[t]
\centering
\includegraphics[width = 5.1in]{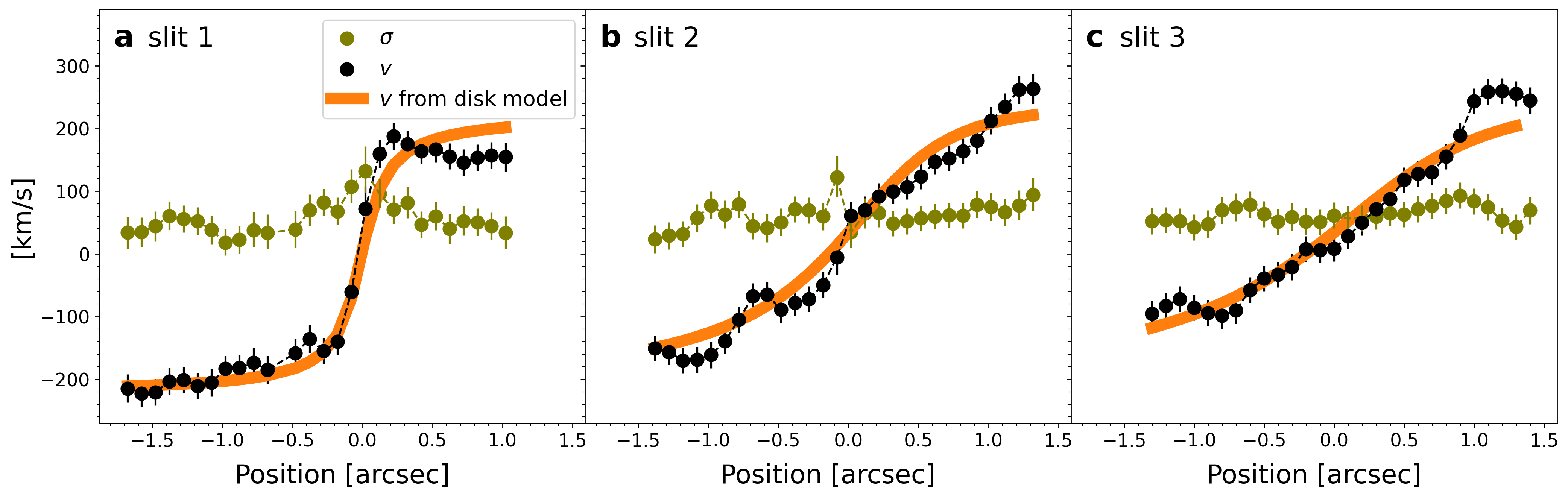}
\caption{ {\bf The rotational velocity along the line of sight ($v$) and velocity dispersion ($\sigma$) profiles measured from three NIRSpec slits, along with predictions of our rotating disk model.} The X axis of each panel indicates the spatial location along a slit where the $v$ and $\sigma$ are measured. Although deviations exist, the disk model captures the bulk trends of the observed $v$ profiles remarkably well, justifying that the ionized gas of the galaxy is predominantly in the ordered motion expected for a disk. Errorbars represent 1-$\sigma$ uncertainties. \label{fig:efig3}}
\end{figure}

One-dimensional H$\alpha$ kinematical profiles have been obtained from the 2D spectra from each pixel row, which spans 0.1\,arcsec along the slit, after convolving them with the instrument line spread function. The 1D spectra were then fit with  a Gaussian using {\sc emcee}.$^{18}$
Outputs of the fits include the rotation velocity (Gaussian centroid) and velocity dispersion (Gaussian width) as a function of the spatial location along the slit, which are shown in Extended Data Fig.~\ref{fig:efig3}. Using the rotation velocities measured from the three slits, we created a velocity map of the galaxy (Fig.~2g).

For the disk modeling, we adopted a flat rotation curve using the {\sc galpy} python package \cite{Bovy2015}. The rotation curve model has a form corresponding to the commonly-used pseudo-isothermal potential\cite{Burkert1995}. It has two free parameters determining the location where the rotation curve turns flat and the plateau velocity, respectively. We created a model galaxy disk using this rotation curve, with an inclination angle and orientation angle of the disk major axis as two more free parameters. Two other free parameters are included which anchor the galaxy center (X, Y). Then, we created a velocity map from this six-parameter disk model by calculating the velocity along the line of sight for each part of the disk. The model map was fitted against the observed velocity map using {\sc emcee}.  After the fitting is done, we calculated the 50th-percentiles of the parameter posterior distributions and used these parameters to create the best-fit disk model. 

Based on the measurements and modeling, we derived the following galaxy kinematic properties. First, we calculated the $v_\mathrm{rot}$, defined as the maximum rotation velocity after inclination correction at 1 half-light radius, where the rotation curve flattens. We calculated the 16-84th percentiles of the $v_\mathrm{rot}$ posterior as the 1-sigma uncertainty from fitting. In addition, wiggles and other small variations from perfect rotating-disk motion are seen in the velocity profiles in Extended Data Fig.~\ref{fig:efig3}. Such features are comparable to the ones commonly seen in other disk galaxies in the local universe, including apparently isolated sources (e.g., [\citen{Rubin1978,Noordermeer2007,deBlok2008,NestorShachar2023}]), and they might also contribute to the uncertainty as a systematic term. To quantify this, we calculated the mean absolute deviation of the observed rotation curve with respect to the model (de-projected for inclination) 
and use it as an upper limit for the systematic uncertainty, which was then combined with the fitting uncertainty as the full uncertainty of $v_\mathrm{rot}$. Second, we measured the averaged intrinsic velocity dispersion $\sigma_\mathrm{int}$ as the median of the dispersion values measured from the two off-center slits in Extended Data Fig.~\ref{fig:efig3}b-c, where the beam-smearing effect is expected to be minimal (see below). The 16th and 84th percentiles of the dispersion values in Extended Data Fig.~\ref{fig:efig3}b-c are quoted as the 1-sigma uncertainty range. The values of $v_\mathrm{rot}$ and $\sigma_\mathrm{int}$ are listed in Tab.~1. Finally, we calculated the circular velocity $v_\mathrm{circ}$, which is used in Fig.~3d, using the following relation adopted in [7, \citen{Ubler2017}]: $v_\mathrm{circ}^2 = v_\mathrm{rot}^2 + 2 \times 1.68\ (R/R_\mathrm{eff})\ \sigma_\mathrm{int}^2$, in which $R$ is where $v_\mathrm{rot}$ was measured in terms of radial distance and equal to one effective radius ($R_\mathrm{eff}$) for the Big Wheel galaxy. The calculated value of $v_\mathrm{circ}$ is $304^{+32}_{-31}\,$km/s.

We investigated the effect of the finite spatial resolution of the observations to the modeling described above, namely the beam-smearing effect \cite{Weiner2006}. We found that it has no substantial impact on the kinematic measurements, since the NIRSpec spatial resolution element (0.13\,arcsec) is substantially smaller than the galaxy size. To justify this, we created a model galaxy with an exponential 2D flux density profile and an intrinsic velocity map identical to what is shown in Fig.~2h. 
We created a 3D mock cube of the intrinsic H$\alpha$ line emission based on this setup. Next, we convolved each wavelength slice of this cube by the instrument PSF to get the synthetic cube ``seen'' at the JWST resolution and produce the corresponding velocity map.
We found that this map deviates from the intrinsic velocity map by no more than 15\,km/s, except for the region within 0.2\,arcsec of the center, confirming that the beam-smearing effect has negligible impact for most parts of the galaxy.

\begin{figure}[t]
\centering
\includegraphics[width = 4 in]{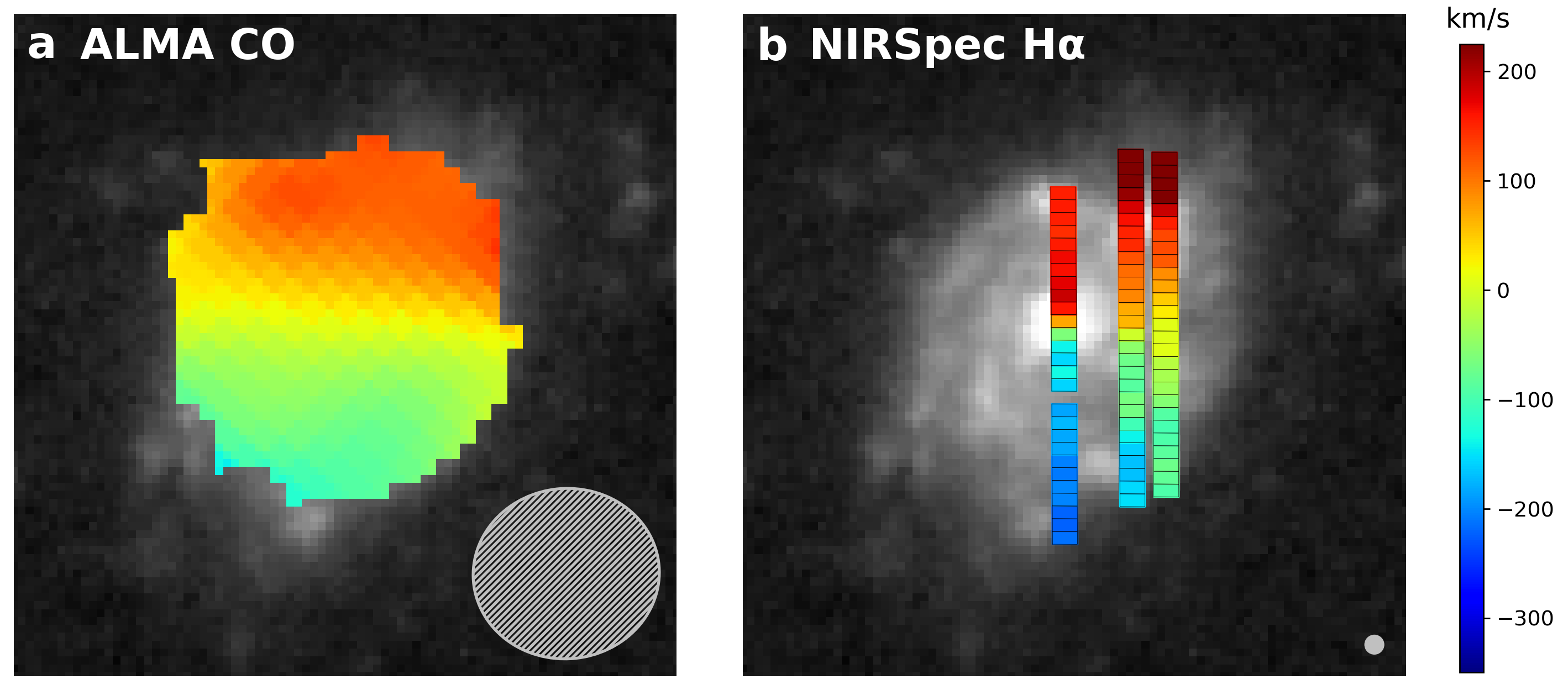}
\caption{ {\bf The velocity map of the Big Wheel galaxy measured from the CO line (a) and the H$\alpha$ line (b).} The values of the line-of-sight velocities are indicated by the colorbar, and the PSFs (in FWHM) of the data are indicated at the lower right corners of the two panels. The CO map has a substantially larger PSF than the H$\alpha$ map and, as a result, shows narrower spans of velocity due to the beam-smearing effect \cite{Weiner2006} Nevertheless, the CO map shows smooth velocity gradients oriented along the galaxy major axis similarly to the H$\alpha$ map, providing additional evidence that the galaxy is consistent with a rotating disk. \label{fig:fig3}}
\end{figure}

\begin{figure}[t]
\centering 
\includegraphics[width = 4.25in]{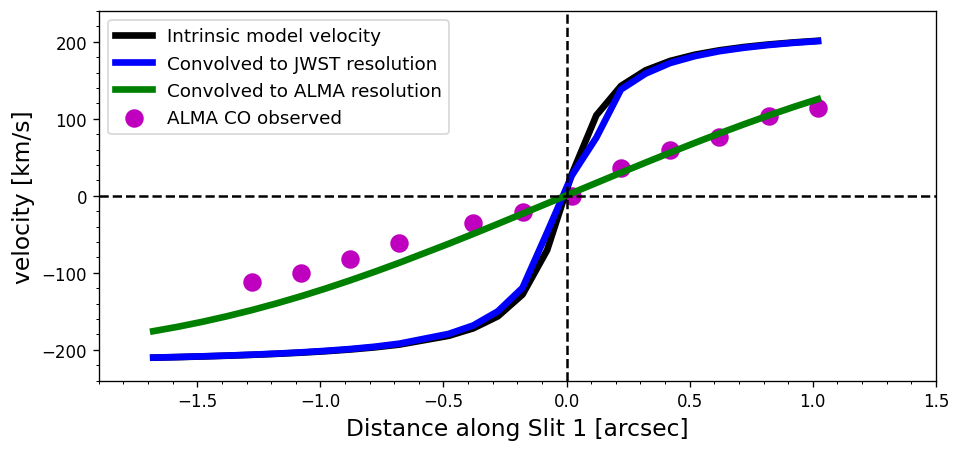}
\caption{ {\bf Galaxy velocity profiles extracted from the 3D emission line models, which are created to examine the beam-smearing effects and compare with observations.} The beam-smearing effect of JWST is found to have negligible impacts on the H$\alpha$ velocity measurements thanks to the exquisite spatial resolution, as demonstrated by that the H$\alpha$ velocity profile ``seen'' at the JWST resolution (blue line) almost fully overlaps with the intrinsic profile (black line). Second, the H$\alpha$ kinematics is found in overall agreement with the CO kinematics. This is demonstrated by that the H$\alpha$ velocity profile ``seen'' at the ALMA resolution (green line) matches the observed ALMA CO velocity profile (purple points). Refer to \emph{Methods} for more details.}
\label{fig:efig7}
\end{figure}

The CO velocity map of the Big Wheel galaxy from the ALMA observations was obtained through the first-moments of the CO(4--3) profiles. The map is shown in Extended Data Fig.~\ref{fig:fig3}a, whereas the H$\alpha$ velocity map is shown in Extended Data Fig.~\ref{fig:fig3}b for comparison. The CO map has a substantially larger PSF than the H$\alpha$ map and, as a result, shows narrower spans of velocity due to the stronger beam-smearing effect \cite{Weiner2006}. Nevertheless, the CO map shows smooth velocity gradients oriented along the galaxy major axis similarly to the H$\alpha$, providing evidence that both the cold molecular and warm ionized gas rotates in a similar fashion. The observed CO kinematics is quantitatively consistent with the H$\alpha$ kinematics once the ALMA PSF is taken into account, as shown in Extended Data Fig.~\ref{fig:efig7}.
In addition, the gas mass was also measured from the ALMA data and listed in Tab.~1, obtained using the CO-to-H$_2$ conversion factors described in [14].

\bmhead{Stellar mass and star formation rate measurements}

\begin{figure}
\centering
\includegraphics[width = 4.95in]{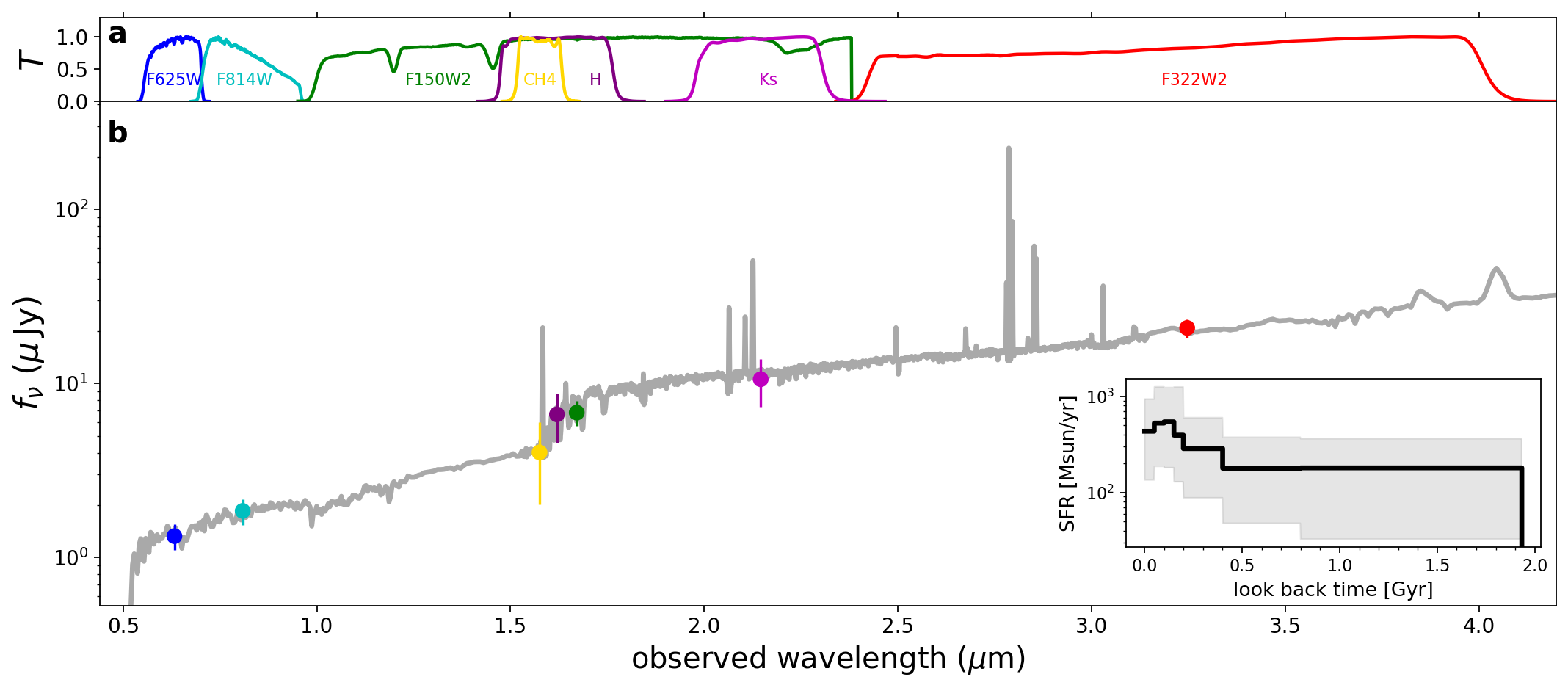}
\caption{ {\bf The observed SED of the Big Wheel along with its best-fit model obtained with the {\sc prospector} code, from which the stellar mass and SFR are inferred for this work.} {\bf (a)} Transmission curves of the photometric filters used in the observations are plotted as a function of wavelength. Each curve is normalized to its maximum. {\bf (b)} The best-fit model (gray curve) is plotted along with the observed flux densities (points with 3-$\sigma$ errorbars). For the observed flux densities, the value measured in a given filter is colored in the same way as in {\bf a}. The vertical axis indicates the flux density. The best-fit star formation history is shown as the black line in the inset, and the gray shade indicates uncertainties. \label{fig:efig4}}
\end{figure}

The stellar mass and star formation rate (SFR) were measured from SED fits with the {\sc{prospector}} tool.$^{17,}$\cite{Johnson2021} Photometric flux densities measured from seven filters are used in the fit: ACS F625W, ACS F814W, NIRCam F150W2, HAWKI CH4, HAWKI H, HAWKI Ks, and NIRCam F322W2. The measured flux densities and the corresponding best-fit SED model are presented in Extended Data Fig.~\ref{fig:efig4}. For the fitting, we fixed the redshift to the value measured from the NIRSpec spectrum. A non-parametric form of the star formation history with a prior favoring continuity\cite{Wang2024} was adopted, binned into seven consecutive ranges in look-back time: 0-50\,Myr, 50-100\,Myr, 100-150\,Myr, 150-200\,Myr, 200-400\,Myr, 400-800\,Myr, and 800-1900\,Myr. A Chabrier\cite{Chabrier2003} initial mass function (IMF) and a flexible dust attenuation law as described in [\citen{Charlot2000}] are adopted. Other assumptions of the fitting procedure, including the forms of the Bayesian priors, are the same as described in [\citen{Wang2024}]. Emission from the AGN was also modeled following the prescription described in [\citen{Wang2024}], although the relevant parameters were unconstrained by the fit. The total stellar mass and SFR (averaged over 100\,Myr) of the galaxy were calculated as the 50th-percentiles of the Bayesian posteriors. The uncertainties were calculated from the 5-95th percentiles of the posteriors and added in quadrature with the systematic uncertainties due to the fitting assumptions as described in [\citen{Pacifici2023}]. The inferred stellar mass and SFR are $3.7^{+2.6}_{-2.2} \times 10^{11}\,M_\odot$ and $2.5^{+7.5}_{-2.1} \times 10^2 \,M_\odot/$yr, respectively.
For comparison, from the H$\alpha$ fluxes contained with the three NIRSpec slits, an SFR of about $1.5\times 10^2 \,M_\odot/$yr (prior to dust correction) was estimated. We noticed though that the slits only cover about 20\% of the galaxy. 

We examined systematic uncertainties of the stellar mass due to the presence of the faint AGN in the center. First, we put an upper limit to its potential contribution to the integrated galaxy fluxes via aperture photometry. Specifically, we measured the fluxes in the HST and JWST filters from within 0.3\,arcsec of the galaxy center, where most of the AGN emission is expected according to the PSFs.  We found that this region contributes less than 1\% and 15\% of the total galaxy light in the HST and JWST filters, respectively, justifying that the AGN has only minor contributions in the rest-frame UV-to-optical.  Second, we performed annulus-by-annulus spatially resolved SED fitting with {\sc prospector} using the HST and JWST data. With identical fitting setups as described above and conservatively excluding the $r<0.3$ arcsec region, we found that the resulting mass decreases by only 0.08 dex with respect to the value from the integrated fitting above  (Extended Data Fig.~\ref{fig:efig2b}c,d). Such a decrease is negligible compared to the stellar mass uncertainty (0.3 dex) reported above, and thus we conclude that the AGN has no significant impact on mass measurements. The mass profile in Extended Data Fig.~\ref{fig:efig2b}c appears to have small variations from a perfectly smooth disk profile (dashed line) by $\pm$0.2 dex, which are, however, within the errorbars. These features are also commonly found with comparable amplitudes in other disk galaxies at $z=0$, including apparently isolated sources\cite{deBlok2008}.

\begin{figure}
\centering
\includegraphics[width = 5.05 in]{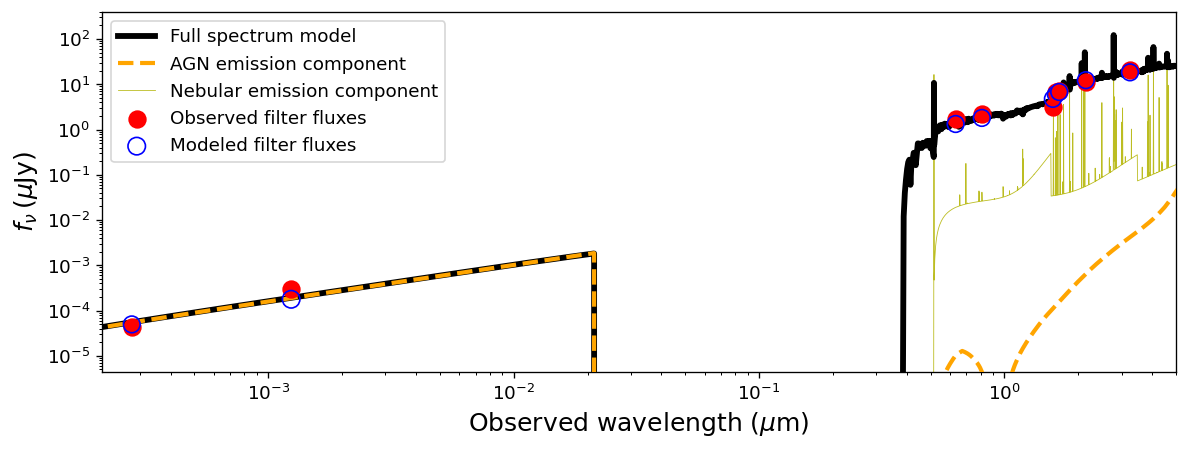}
\caption{ {\bf The observed SED of the Big Wheel, along with its best-fit models obtained with the CIGALE code.} The best-fit fluxes and corresponding full spectrum model are shown as open circles and the thick black curve, respectively, whereas individual components of the model spectrum are shown as curves in other colors. The code models and fits the X-ray fluxes, as well as the optical and infrared ones. The best-fit AGN emission component (orange dashed curve) contributes to only $\lesssim 10^{-3}$ of the observed fluxes in the optical and infrared (filled circles). \label{fig:efig5}}
\end{figure}

We conducted two consistency checks for the SED fit described above. First, we repeated the fit with a parametric star formation history, which was adopted for stellar mass measurements by the majority of the observational studies included for comparison in Fig.~3. The adopted history in a delayed exponentially declining form. The stellar mass inferred is $1.7^{+1.4}_{-0.7}\times 10^{11}\,M_\odot$, moderately lower than that reported from the non-parametric fit but still consistent within uncertainties. Furthermore, we also performed SED fit using the code {\sc CIGALE} \cite{Burgarella2005,Noll2009,Boquien2019}. It  models both optical/infrared and X-ray flux densities, the latter measured with Chandra, based on the prescriptions described by [\citen{Yang2022}]. The fitting assumptions are similar to those adopted for {\sc prospector}, except that the SFH is assumed to have a delayed exponentially declining form. According to the fitting results in Extended Data Fig.~\ref{fig:efig5}, the AGN emission (orange curve) has only minimal contribution ($\lesssim10^{-3}$) to the observed fluxes (filled circles) in the optical and infrared filters. The best-fit stellar mass and SFR (averaged over 100\,Myr) are $2.5 \times 10^{11}\,M_\odot$ and $1.5 \times 10^2 \,M_\odot/$yr, respectively, both within the uncertainty ranges inferred by {\sc prospector}.

\bmhead{Comparison sample of disk galaxies from the literature}

Disk galaxies at $z$=3-4 from the literature that were confirmed by kinematics and measured for size are included in Fig.~3a. 
They are from ground-based studies based on rest-frame optical emission lines \cite{Gnerucci2011, Livermore2015,Turner2017,Gillman2019}, except for two objects which are based on the CO emission lines with ALMA \cite{Swinbank2015} and optical emission lines with JWST \cite{Wu2023}.  All the disk galaxies were selected to have rotational velocities greater than the velocity dispersions, i.e., $v_\mathrm{rot}>\sigma_\mathrm{int}$, a lenient disk criterion commonly adopted in the literature.$^{7,19,}$\cite{Simons2017,Wisnioski2015} The stellar masses of all the disk galaxies were inferred using a Chabrier\cite{Chabrier2003} or Kroupa IMF\cite{Kroupa2001}, except for the AMAZE\cite{Maiolino2008} subsample in Gnerucci et al\cite{Gnerucci2011} which adopt a Salpeter IMF\cite{Salpeter1955}. We adjust the latter to the Chabrier IMF by subtracting the reported mass values by 0.24 dex \cite{Courteau2014,Santini2015}. We also note that disks have also been identified kinematically beyond z=4, which are not included in Fig.~3. These kinematically-confirmed disks were mostly discovered with ALMA, with one discovered with JWST.$^{4,}$\cite{Carniani2013,Jones2017, Tadaki2018,Neeleman2019,Rizzo2021,Rizzo2023,Fraternali2021,Fujimoto2021, Lelli2021,Tsukui2021,Herrera-Camus2022,Nelson2023,Parlanti2023,Roman-Oliveira2023} The largest galaxies reported by these studies are around 5\,kiloparsecs in radius \cite{Neeleman2019, Parlanti2023}, which are approximately two times smaller than the Big Wheel galaxy. 

We note that kinematical measurements are essential for the confirmation of disks, because measurements using imaging data alone may be subject to ambiguity of interpretation due to clumps, mergers, dust, and peculiar galaxy geometry, all found common at $z\gtrsim 1$.$^{7,}$\cite{vanderwel2014,Guo2015,Vega-Ferrero2024} As an evidence supporting this argument, we examined the galaxies from [5] at $z=3-4$ with large radii ($>0.5\,$arcsec) measured from HST near-infrared images, and found that the majority of them, if not all, have irregular or major merger-like morphology rather than disk-like. For this reason, only galaxies with kinematic measurements were considered for comparison in Fig.~3a.

\bmhead{Other measurements from the literature used for comparison} 

The size-mass scaling relations in Fig.~3b are measured at or around the rest-frame $0.5\,\mu$m, the same wavelength where the size of the Big Wheel galaxy is measured. We include only star-forming galaxies.$^{8,9,11,12,}$\cite{Mowla2019,Ormerod2024} The size-mass relation of Ward et al[8] are measured in two redshift bins, $z=2-3$ and $z=3-5.5$, and we adopted the average values to match the redshift of our galaxy ($z=3.245$). The size-mass relation of Costantin et al$^{11}$ is provided as an analytical function of redshift and we adopted the relation at $z=3.2$. The size-mass relation of Mowla et al\cite{Mowla2019} is determined in the work as the median trend of a statistical galaxy sample at $z=0.1-0.5$.  

Based on the literature measurements, the probability of finding a star forming galaxy, independent of kinematical confirmation, with the size of the Big Wheel at its stellar mass and redshift in random fields was obtained as follows. First, we calculated the expected galaxy size at the mass and redshift of the Big Wheel, using the relations from two most recent and complete JWST studies as shown in Fig.~3b$^{8,9}$. We found it to be 0.41 dex and 0.70 dex smaller than the observed size of the Big Wheel in log using [8] and [9] relations, respectively. The distribution of sizes is well constrained as a log-normal in these studies with 
an intrinsic log-normal scatter in exponential units ($\sigma_{\mathrm{log(R_e)}}$) of these size-mass relations of 0.22 dex and 0.28 dex, respectively. Therefore
the corresponding probabilities of finding a galaxy equal to or larger than the Big Wheel in random fields are 2\% and 0.6\% according to the two studies, respectively.

In addition, star forming galaxies with masses larger than 10$^{11}\ M_\odot$, such as the Big Wheel (which has a mass of at least $10^{11}\ M_\odot$), are extremely rare, independent of their sizes. This can be quantified again
by using the previously quoted JWST results$^{8,9}$.
Only two star forming galaxies 
above a mass of $10^{11}\ M_\odot$ are detected in these surveys above redshift 3 over a combined area of 440 arcmin$^2$ and no galaxies have a mass above $10^{11}\ M_\odot$, the lower limit for the Big Wheel considering different methodologies.
Thus, in order to randomly discover a galaxy with a mass of at least $10^{11}\ M_\odot$ at redshift greater than 3 and sizes similar to the Big Wheel (using the conservative probability limit of 2\%), if the environment does not play a role, we would have required a survey with a size of at least 11 thousand arcmin$^2$.

Regarding the general galaxy population shown in Fig.~3c, they are selected to be at $z=3.0-3.5$ from [\citen{Santini2015,Barro2019}]. In the same panel, scaling relations from two other literature studies are also shown \cite{Speagle2014,Popesso2023}. The $z\sim 3.5$ galaxies in Fig.~3d were selected to have the rotational velocity ($v_\mathrm{rot}$) greater than the velocity dispersion ($\sigma_\mathrm{int}$). We inferred their circular velocities ($v_\mathrm{circ}$) by performing a correction for disordered motions to $v_\mathrm{rot}$, which was measured at 2 times the half-light radius \cite{Turner2017}, following the same equation as used for the Big Wheel above. The $z\sim 0$ galaxies in the same panel are from [22] and [\citen{Lelli2016}]. The stellar masses in these studies were all inferred assuming a Chabrier\cite{Chabrier2003} IMF.

\backmatter

\noindent

\backmatter
\bmhead{Supplementary information}

Supplementary Information is available for this paper.

\bmhead{Acknowledgments}

This project was supported by the European Research Council (ERC) Consolidator Grant 864361 (CosmicWeb). A.P. acknowledges the support from Fondazione Cariplo grant no. 2020-0902. M.M. acknowledges funding from NASA via HST-GO-17065. T.N. acknowledge support from Australian Research Council Laureate Fellowship FL180100060.
This work is based in part on observations made with the NASA/ESA/CSA James Webb Space Telescope. The data were obtained from the Mikulski Archive for Space Telescopes at the Space Telescope Science Institute, which is operated by the Association of Universities for Research in Astronomy, Inc., under NASA contract NAS 5-03127 for JWST. These observations are associated with program \#1835.
Support for program \#1835 was provided by NASA through a grant from the Space Telescope Science Institute, which is operated by the Association of Universities for Research in Astronomy, Inc., under NASA contract NAS 5-03127. This research is based on observations made with the NASA/ESA Hubble Space Telescope obtained from the Space Telescope Science Institute, which is operated by the Association of Universities for Research in Astronomy, Inc., under NASA contract NAS 5–26555. These observations are associated with program 17065.
ALMA is a partnership of ESO (representing its member states), NSF (USA) and NINS (Japan), together with NRC (Canada), MOST and ASIAA (Taiwan), and KASI (Republic of Korea), in cooperation with the Republic of Chile. The Joint ALMA Observatory is operated by ESO, AUI/NRAO and NAOJ. The scientific results reported in this article are based in part on observations made by the Chandra X-ray Observatory. This work is also based on observations collected at the European Southern Observatory under ESO programme (110.23ZX).

\bmhead{Author contributions} 
W.W. was responsible for the JWST and HST observations, reduced the data from JWST, HST, VLT, conducted the imaging, spectroscopic, and modeling analyses of this work, produced the figures and worked in collaboration with S.C. to the text writing. S.C. led the proposal and planning for the JWST, HST, and VLT observations used in this work, contributed to the observational, theoretical analyses and interpretation of the data and results including text writing. A.P. was responsible for the ALMA observations, conducted the analysis and modeling of ALMA data, contributed to the interpretation and discussion. M.G. contributed to the observation planning, assisted in reducing the VLT data, performed the SED fitting, and contributed to the interpretation of the results. A.T. planned the Chandra observations, reduced the Chandra data, and contributed to analyses relevant to the AGN properties. C.S. assisted with the JWST target selection, observation planning, and data reduction, and contributed to the interpretation of science results. 
M.Fossati and M.Fumagalli assisted with the JWST proposal preparation, contributed to the interpretation and discussion of the observational results. T.L., S.dB., and G.Q. contributed to the comparison with simulations and relevant discussion. M.M. assisted with the JWST NIRSpec MSA mask design, preparation of the APT files, advised on observing and data reduction strategies, and contributed to the interpretation and discussion. S.G., J.M., R.M., T.N., and G.P. contributed to the proposal planning, data interpretation and discussion for the JWST and HST observations.

\bmhead{Data and code availability}
The JWST NIRCam data (Program ID: GO 1835) are publicly available at \url{https://mast.stsci.edu/search/ui/#/jwst}. The HST data (Program ID: 17065) are publicly available at \url{https://mast.stsci.edu/search/ui/#/hst}. This paper also makes use of the ALMA dataset ADS/JAO.ALMA\#2021.1.00793.S. which can be obtained at \url{http://almascience.nao.ac.jp}. 
The {\sc Source Extractor} code is available at \url{https://astromatic.github.io/sextractor/}. The {\sc statmorph} package is available to download at \url{https://github.com/vrodgom/statmorph/}.  The {\sc emcee} package is available at \url{https://emcee.readthedocs.io/en/stable/}. The {\sc galpy} package is available at \url{https://github.com/jobovy/galpy}.  The {\sc prospector} code is available to download at \url{https://github.com/bd-j/prospector}. The {\sc CIGALE} code is available at \url{https://cigale.lam.fr/}. The {\sc drizzle} code is available to download at \url{https://github.com/spacetelescope/drizzle}.

\bmhead{Competing interests} The authors declare no competing interests.

\bmhead{Corresponding author} Correspondence and requests for materials should be addressed to Weichen Wang (weichen.wang@unimib.it).

\bibliography{bigwheel}

\begin{thebibliography}{10}
\expandafter\ifx\csname url\endcsname\relax
  \def\url#1{\burl{#1}}\fi
\expandafter\ifx\csname urlprefix\endcsname\relax\def\urlprefix{URL }\fi
\providecommand{\bibinfo}[2]{#2}
\providecommand{\eprint}[2][]{\url{#2}}
\providecommand{\doi}[1]{\url{https://doi.org/#1}}
\bibcommenthead

\bibitem{Genzel2006}
\bibinfo{author}{{Genzel}, R.} \emph{et~al.}
\newblock \bibinfo{title}{{The rapid formation of a large rotating disk galaxy three billion years after the Big Bang}}.
\newblock \emph{\bibinfo{journal}{Nature}} \textbf{\bibinfo{volume}{442}}, \bibinfo{pages}{786--789} (\bibinfo{year}{2006}).

\bibitem{Law2012}
\bibinfo{author}{{Law}, D.~R.} \emph{et~al.}
\newblock \bibinfo{title}{{High velocity dispersion in a rare grand-design spiral galaxy at redshift z = 2.18}}.
\newblock \emph{\bibinfo{journal}{Nature}} \textbf{\bibinfo{volume}{487}}, \bibinfo{pages}{338--340} (\bibinfo{year}{2012}).

\bibitem{Genzel2017}
\bibinfo{author}{{Genzel}, R.} \emph{et~al.}
\newblock \bibinfo{title}{{Strongly baryon-dominated disk galaxies at the peak of galaxy formation ten billion years ago}}.
\newblock \emph{\bibinfo{journal}{Nature}} \textbf{\bibinfo{volume}{543}}, \bibinfo{pages}{397--401} (\bibinfo{year}{2017}).

\bibitem{Rizzo2020}
\bibinfo{author}{{Rizzo}, F.} \emph{et~al.}
\newblock \bibinfo{title}{{A dynamically cold disk galaxy in the early Universe}}.
\newblock \emph{\bibinfo{journal}{Nature}} \textbf{\bibinfo{volume}{584}}, \bibinfo{pages}{201--204} (\bibinfo{year}{2020}).

\bibitem{vanderwel2014b}
\bibinfo{author}{{van der Wel}, A.} \emph{et~al.}
\newblock \bibinfo{title}{{3D-HST+CANDELS: The Evolution of the Galaxy Size-Mass Distribution since z = 3}}.
\newblock \emph{\bibinfo{journal}{Astrophys. J.}} \textbf{\bibinfo{volume}{788}}, \bibinfo{pages}{28} (\bibinfo{year}{2014}).

\bibitem{Ogle2016}
\bibinfo{author}{{Ogle}, P.~M.}, \bibinfo{author}{{Lanz}, L.}, \bibinfo{author}{{Nader}, C.} \& \bibinfo{author}{{Helou}, G.}
\newblock \bibinfo{title}{{Superluminous Spiral Galaxies}}.
\newblock \emph{\bibinfo{journal}{Astrophys. J.}} \textbf{\bibinfo{volume}{817}}, \bibinfo{pages}{109} (\bibinfo{year}{2016}).

\bibitem{ForsterSchreiber2020}
\bibinfo{author}{{F{\"o}rster Schreiber}, N.~M.} \& \bibinfo{author}{{Wuyts}, S.}
\newblock \bibinfo{title}{{Star-Forming Galaxies at Cosmic Noon}}.
\newblock \emph{\bibinfo{journal}{Annu. Rev. Astron. Astrophys.}} \textbf{\bibinfo{volume}{58}}, \bibinfo{pages}{661--725} (\bibinfo{year}{2020}).

\bibitem{Ward2023}
\bibinfo{author}{{Ward}, E.} \emph{et~al.}
\newblock \bibinfo{title}{{Evolution of the Size{\textendash}Mass Relation of Star-forming Galaxies Since z = 5.5 Revealed by CEERS}}.
\newblock \emph{\bibinfo{journal}{Astrophys. J.}} \textbf{\bibinfo{volume}{962}}, \bibinfo{pages}{176} (\bibinfo{year}{2024}).

\bibitem{Varadaraj2024}
\bibinfo{author}{{Varadaraj}, R.~G.} \emph{et~al.}
\newblock \bibinfo{title}{{The sizes of bright Lyman-break galaxies at z = 3-5 with JWST PRIMER}}.
\newblock \emph{\bibinfo{journal}{Mon. Not. R. Astron.}}  (\bibinfo{year}{2024}).

\bibitem{MMW1998}
\bibinfo{author}{{Mo}, H.~J.}, \bibinfo{author}{{Mao}, S.} \& \bibinfo{author}{{White}, S. D.~M.}
\newblock \bibinfo{title}{{The formation of galactic discs}}.
\newblock \emph{\bibinfo{journal}{Mon. Not. R. Astron.}} \textbf{\bibinfo{volume}{295}}, \bibinfo{pages}{319--336} (\bibinfo{year}{1998}).

\bibitem{Costantin2023}
\bibinfo{author}{{Costantin}, L.} \emph{et~al.}
\newblock \bibinfo{title}{{Expectations of the Size Evolution of Massive Galaxies at 3 {\ensuremath{\leq}} z {\ensuremath{\leq}} 6 from the TNG50 Simulation: The CEERS/JWST View}}.
\newblock \emph{\bibinfo{journal}{Astrophys. J.}} \textbf{\bibinfo{volume}{946}}, \bibinfo{pages}{71} (\bibinfo{year}{2023}).

\bibitem{LaChance2024}
\bibinfo{author}{{LaChance}, P.} \emph{et~al.}
\newblock \bibinfo{title}{{The evolution of galaxy morphology from redshift z=6 to 3: Mock JWST observations of galaxies in the ASTRID simulation}}.
\newblock \emph{\bibinfo{journal}{Preprint at arxiv.org/abs/2401.16608}}  (\bibinfo{year}{2024}).

\bibitem{Lelli2016}
\bibinfo{author}{{Lelli}, F.}, \bibinfo{author}{{McGaugh}, S.~S.} \& \bibinfo{author}{{Schombert}, J.~M.}
\newblock \bibinfo{title}{{SPARC: Mass Models for 175 Disk Galaxies with Spitzer Photometry and Accurate Rotation Curves}}.
\newblock \emph{\bibinfo{journal}{Astron. J.}} \textbf{\bibinfo{volume}{152}}, \bibinfo{pages}{157} (\bibinfo{year}{2016}).

\bibitem{Pensabene2024}
\bibinfo{author}{{Pensabene}, A.} \emph{et~al.}
\newblock \bibinfo{title}{{ALMA survey of a massive node of the Cosmic Web at z {\ensuremath{\sim}} 3. I. Discovery of a large overdensity of CO emitters}}.
\newblock \emph{\bibinfo{journal}{Astron. Astrophys.}} \textbf{\bibinfo{volume}{684}}, \bibinfo{pages}{A119} (\bibinfo{year}{2024}).

\bibitem{GildePaz2007}
\bibinfo{author}{{Gil de Paz}, A.} \emph{et~al.}
\newblock \bibinfo{title}{{The GALEX Ultraviolet Atlas of Nearby Galaxies}}.
\newblock \emph{\bibinfo{journal}{Astrophys. J. Suppl. Ser.}} \textbf{\bibinfo{volume}{173}}, \bibinfo{pages}{185--255} (\bibinfo{year}{2007}).

\bibitem{Rodriguez-Gomez2019}
\bibinfo{author}{{Rodriguez-Gomez}, V.} \emph{et~al.}
\newblock \bibinfo{title}{{The optical morphologies of galaxies in the IllustrisTNG simulation: a comparison to Pan-STARRS observations}}.
\newblock \emph{\bibinfo{journal}{Mon. Not. R. Astron.}} \textbf{\bibinfo{volume}{483}}, \bibinfo{pages}{4140--4159} (\bibinfo{year}{2019}).

\bibitem{Foreman-Mackey2013}
\bibinfo{author}{{Foreman-Mackey}, D.}, \bibinfo{author}{{Hogg}, D.~W.}, \bibinfo{author}{{Lang}, D.} \& \bibinfo{author}{{Goodman}, J.}
\newblock \bibinfo{title}{{emcee: The MCMC Hammer}}.
\newblock \emph{\bibinfo{journal}{Publ. Astron. Soc. Pac.}} \textbf{\bibinfo{volume}{125}}, \bibinfo{pages}{306} (\bibinfo{year}{2013}).

\bibitem{Kassin2012}
\bibinfo{author}{{Kassin}, S.~A.} \emph{et~al.}
\newblock \bibinfo{title}{{The Epoch of Disk Settling: z \raisebox{-0.5ex}\textasciitilde 1 to Now}}.
\newblock \emph{\bibinfo{journal}{Astrophys. J.}} \textbf{\bibinfo{volume}{758}}, \bibinfo{pages}{106} (\bibinfo{year}{2012}).

\bibitem{Ubler2017}
\bibinfo{author}{{{\"U}bler}, H.} \emph{et~al.}
\newblock \bibinfo{title}{{The Evolution of the Tully-Fisher Relation between z {\ensuremath{\sim}} 2.3 and z {\ensuremath{\sim}} 0.9 with KMOS$^{3D}$}}.
\newblock \emph{\bibinfo{journal}{Astrophys. J.}} \textbf{\bibinfo{volume}{842}}, \bibinfo{pages}{121} (\bibinfo{year}{2017}).

\bibitem{Leja2017}
\bibinfo{author}{{Leja}, J.}, \bibinfo{author}{{Johnson}, B.~D.}, \bibinfo{author}{{Conroy}, C.}, \bibinfo{author}{{van Dokkum}, P.~G.} \& \bibinfo{author}{{Byler}, N.}
\newblock \bibinfo{title}{{Deriving Physical Properties from Broadband Photometry with Prospector: Description of the Model and a Demonstration of its Accuracy Using 129 Galaxies in the Local Universe}}.
\newblock \emph{\bibinfo{journal}{Astrophys. J.}} \textbf{\bibinfo{volume}{837}}, \bibinfo{pages}{170} (\bibinfo{year}{2017}).

\bibitem{Nanayakkara2024}
\bibinfo{author}{{Nanayakkara}, T.} \emph{et~al.}
\newblock \bibinfo{title}{{A population of faint, old, and massive quiescent galaxies at $3 <z <4$ revealed by JWST NIRSpec Spectroscopy}}.
\newblock \emph{\bibinfo{journal}{Scientific Reports}} \textbf{\bibinfo{volume}{14}}, \bibinfo{pages}{3724} (\bibinfo{year}{2024}).

\bibitem{DiTeodoro2021}
\bibinfo{author}{{Di Teodoro}, E.~M.}, \bibinfo{author}{{Posti}, L.}, \bibinfo{author}{{Ogle}, P.~M.}, \bibinfo{author}{{Fall}, S.~M.} \& \bibinfo{author}{{Jarrett}, T.}
\newblock \bibinfo{title}{{Rotation curves and scaling relations of extremely massive spiral galaxies}}.
\newblock \emph{\bibinfo{journal}{Mon. Not. R. Astron.}} \textbf{\bibinfo{volume}{507}}, \bibinfo{pages}{5820--5831} (\bibinfo{year}{2021}).

\bibitem{Bullock2001}
\bibinfo{author}{{Bullock}, J.~S.} \emph{et~al.}
\newblock \bibinfo{title}{{A Universal Angular Momentum Profile for Galactic Halos}}.
\newblock \emph{\bibinfo{journal}{Astrophys. J.}} \textbf{\bibinfo{volume}{555}}, \bibinfo{pages}{240--257} (\bibinfo{year}{2001}).

\bibitem{Bryan2013}
\bibinfo{author}{{Bryan}, S.~E.} \emph{et~al.}
\newblock \bibinfo{title}{{The impact of baryons on the spins and shapes of dark matter haloes}}.
\newblock \emph{\bibinfo{journal}{Mon. Not. R. Astron.}} \textbf{\bibinfo{volume}{429}}, \bibinfo{pages}{3316--3329} (\bibinfo{year}{2013}).

\bibitem{Zjupa2017}
\bibinfo{author}{{Zjupa}, J.} \& \bibinfo{author}{{Springel}, V.}
\newblock \bibinfo{title}{{Angular momentum properties of haloes and their baryon content in the Illustris simulation}}.
\newblock \emph{\bibinfo{journal}{Mon. Not. R. Astron.}} \textbf{\bibinfo{volume}{466}}, \bibinfo{pages}{1625--1647} (\bibinfo{year}{2017}).

\bibitem{Jian2012}
\bibinfo{author}{{Jian}, H.-Y.}, \bibinfo{author}{{Lin}, L.} \& \bibinfo{author}{{Chiueh}, T.}
\newblock \bibinfo{title}{{Environmental Dependence of the Galaxy Merger Rate in a {\ensuremath{\Lambda}}CDM Universe}}.
\newblock \emph{\bibinfo{journal}{Astrophys. J.}} \textbf{\bibinfo{volume}{754}}, \bibinfo{pages}{26} (\bibinfo{year}{2012}).

\bibitem{Governato2009}
\bibinfo{author}{{Governato}, F.} \emph{et~al.}
\newblock \bibinfo{title}{{Forming a large disc galaxy from a $z < 1$ major merger}}.
\newblock \emph{\bibinfo{journal}{Mon. Not. R. Astron.}} \textbf{\bibinfo{volume}{398}}, \bibinfo{pages}{312--320} (\bibinfo{year}{2009}).

\bibitem{Hopkins2009}
\bibinfo{author}{{Hopkins}, P.~F.}, \bibinfo{author}{{Cox}, T.~J.}, \bibinfo{author}{{Younger}, J.~D.} \& \bibinfo{author}{{Hernquist}, L.}
\newblock \bibinfo{title}{{How do Disks Survive Mergers?}}
\newblock \emph{\bibinfo{journal}{Astrophys. J.}} \textbf{\bibinfo{volume}{691}}, \bibinfo{pages}{1168--1201} (\bibinfo{year}{2009}).

\bibitem{Stewart2017}
\bibinfo{author}{{Stewart}, K.~R.} \emph{et~al.}
\newblock \bibinfo{title}{{High Angular Momentum Halo Gas: A Feedback and Code-independent Prediction of LCDM}}.
\newblock \emph{\bibinfo{journal}{Astrophys. J.}} \textbf{\bibinfo{volume}{843}}, \bibinfo{pages}{47} (\bibinfo{year}{2017}).

\bibitem{Bushouse2022}
\bibinfo{author}{{Bushouse}, H.} \emph{et~al.}
\newblock \bibinfo{title}{{JWST Calibration Pipeline}}, \bibinfo{version}{1.8.0} (\bibinfo{year}{2022}).

\bibitem{Bagley2023}
\bibinfo{author}{{Bagley}, M.~B.} \emph{et~al.}
\newblock \bibinfo{title}{{CEERS Epoch 1 NIRCam Imaging: Reduction Methods and Simulations Enabling Early JWST Science Results}}.
\newblock \emph{\bibinfo{journal}{Astrophys. J. Lett.}} \textbf{\bibinfo{volume}{946}}, \bibinfo{pages}{L12} (\bibinfo{year}{2023}).

\bibitem{esorex2015}
\bibinfo{author}{{ESO CPL Development Team}}.
\newblock \bibinfo{title}{{EsoRex: ESO Recipe Execution Tool}}.
\newblock \bibinfo{howpublished}{Astrophysics Source Code Library, record ascl:1504.003} (\bibinfo{year}{2015}).

\bibitem{Bertin1996}
\bibinfo{author}{{Bertin}, E.} \& \bibinfo{author}{{Arnouts}, S.}
\newblock \bibinfo{title}{{SExtractor: Software for source extraction.}}
\newblock \emph{\bibinfo{journal}{A{\&}AS}} \textbf{\bibinfo{volume}{117}}, \bibinfo{pages}{393--404} (\bibinfo{year}{1996}).

\bibitem{Fruchter2002}
\bibinfo{author}{{Fruchter}, A.~S.} \& \bibinfo{author}{{Hook}, R.~N.}
\newblock \bibinfo{title}{{Drizzle: A Method for the Linear Reconstruction of Undersampled Images}}.
\newblock \emph{\bibinfo{journal}{Publ. Astron. Soc. Pac.}} \textbf{\bibinfo{volume}{114}}, \bibinfo{pages}{144--152} (\bibinfo{year}{2002}).

\bibitem{Kewley2001}
\bibinfo{author}{{Kewley}, L.~J.}, \bibinfo{author}{{Dopita}, M.~A.}, \bibinfo{author}{{Sutherland}, R.~S.}, \bibinfo{author}{{Heisler}, C.~A.} \& \bibinfo{author}{{Trevena}, J.}
\newblock \bibinfo{title}{{Theoretical Modeling of Starburst Galaxies}}.
\newblock \emph{\bibinfo{journal}{Astrophys. J.}} \textbf{\bibinfo{volume}{556}}, \bibinfo{pages}{121--140} (\bibinfo{year}{2001}).

\bibitem{Perrin2014}
\bibinfo{author}{{Perrin}, M.~D.} \emph{et~al.}
\newblock \bibinfo{editor}{{Oschmann}, J., Jacobus~M.}, \bibinfo{editor}{{Clampin}, M.}, \bibinfo{editor}{{Fazio}, G.~G.} \& \bibinfo{editor}{{MacEwen}, H.~A.} (eds) \emph{\bibinfo{title}{{Updated point spread function simulations for JWST with WebbPSF}}}.
\newblock (eds \bibinfo{editor}{{Oschmann}, J., Jacobus~M.}, \bibinfo{editor}{{Clampin}, M.}, \bibinfo{editor}{{Fazio}, G.~G.} \& \bibinfo{editor}{{MacEwen}, H.~A.}) \emph{\bibinfo{booktitle}{Space Telescopes and Instrumentation 2014: Optical, Infrared, and Millimeter Wave}}, Vol. \bibinfo{volume}{9143} of \emph{\bibinfo{series}{Society of Photo-Optical Instrumentation Engineers (SPIE) Conference Series}}, \bibinfo{pages}{91433X} (\bibinfo{year}{2014}).

\bibitem{CASA2022}
\bibinfo{author}{{CASA Team}} \emph{et~al.}
\newblock \bibinfo{title}{{CASA, the Common Astronomy Software Applications for Radio Astronomy}}.
\newblock \emph{\bibinfo{journal}{Publ. Astron. Soc. Pac.}} \textbf{\bibinfo{volume}{134}}, \bibinfo{pages}{114501} (\bibinfo{year}{2022}).

\bibitem{Ueda2014}
\bibinfo{author}{{Ueda}, Y.}, \bibinfo{author}{{Akiyama}, M.}, \bibinfo{author}{{Hasinger}, G.}, \bibinfo{author}{{Miyaji}, T.} \& \bibinfo{author}{{Watson}, M.~G.}
\newblock \bibinfo{title}{{Toward the Standard Population Synthesis Model of the X-Ray Background: Evolution of X-Ray Luminosity and Absorption Functions of Active Galactic Nuclei Including Compton-thick Populations}}.
\newblock \emph{\bibinfo{journal}{Astrophys. J.}} \textbf{\bibinfo{volume}{786}}, \bibinfo{pages}{104} (\bibinfo{year}{2014}).

\bibitem{Baldwin1981}
\bibinfo{author}{{Baldwin}, J.~A.}, \bibinfo{author}{{Phillips}, M.~M.} \& \bibinfo{author}{{Terlevich}, R.}
\newblock \bibinfo{title}{{Classification parameters for the emission-line spectra of extragalactic objects.}}
\newblock \emph{\bibinfo{journal}{Publ. Astron. Soc. Pac.}} \textbf{\bibinfo{volume}{93}}, \bibinfo{pages}{5--19} (\bibinfo{year}{1981}).

\bibitem{Sersic1963}
\bibinfo{author}{{S{\'e}rsic}, J.~L.}
\newblock \bibinfo{title}{{Influence of the atmospheric and instrumental dispersion on the brightness distribution in a galaxy}}.
\newblock \emph{\bibinfo{journal}{Bol. Asoc. Argent. Astron. Plata Argent.}} \textbf{\bibinfo{volume}{6}}, \bibinfo{pages}{41--43} (\bibinfo{year}{1963}).

\bibitem{Dominguez2013}
\bibinfo{author}{{Dom{\'\i}nguez}, A.} \emph{et~al.}
\newblock \bibinfo{title}{{Dust Extinction from Balmer Decrements of Star-forming Galaxies at 0.75 $<$= z $<$= 1.5 with Hubble Space Telescope/Wide-Field-Camera 3 Spectroscopy from the WFC3 Infrared Spectroscopic Parallel Survey}}.
\newblock \emph{\bibinfo{journal}{Astrophys. J.}} \textbf{\bibinfo{volume}{763}}, \bibinfo{pages}{145} (\bibinfo{year}{2013}).

\bibitem{DEugenio2024}
\bibinfo{author}{{D'Eugenio}, F.} \emph{et~al.}
\newblock \bibinfo{title}{{JADES Data Release 3 -- NIRSpec/MSA spectroscopy for 4,000 galaxies in the GOODS fields}}.
\newblock \emph{\bibinfo{journal}{Preprint at arxiv.org/abs/2404.06531}}  (\bibinfo{year}{2024}).

\bibitem{Calzetti2001}
\bibinfo{author}{{Calzetti}, D.}
\newblock \bibinfo{title}{{The Dust Opacity of Star-forming Galaxies}}.
\newblock \emph{\bibinfo{journal}{Publ. Astron. Soc. Pac.}} \textbf{\bibinfo{volume}{113}}, \bibinfo{pages}{1449--1485} (\bibinfo{year}{2001}).

\bibitem{Nelson2019}
\bibinfo{author}{{Nelson}, E.~J.} \emph{et~al.}
\newblock \bibinfo{title}{{Millimeter Mapping at z {\ensuremath{\sim}} 1: Dust-obscured Bulge Building and Disk Growth}}.
\newblock \emph{\bibinfo{journal}{Astrophys. J.}} \textbf{\bibinfo{volume}{870}}, \bibinfo{pages}{130} (\bibinfo{year}{2019}).

\bibitem{Mowla2019}
\bibinfo{author}{{Mowla}, L.~A.} \emph{et~al.}
\newblock \bibinfo{title}{{COSMOS-DASH: The Evolution of the Galaxy Size-Mass Relation since z {\ensuremath{\sim}} 3 from New Wide-field WFC3 Imaging Combined with CANDELS/3D-HST}}.
\newblock \emph{\bibinfo{journal}{Astrophys. J.}} \textbf{\bibinfo{volume}{880}}, \bibinfo{pages}{57} (\bibinfo{year}{2019}).

\bibitem{Lange2015}
\bibinfo{author}{{Lange}, R.} \emph{et~al.}
\newblock \bibinfo{title}{{Galaxy And Mass Assembly (GAMA): mass-size relations of z$<$0.1 galaxies subdivided by S{\'e}rsic index, colour and morphology}}.
\newblock \emph{\bibinfo{journal}{Mon. Not. R. Astron.}} \textbf{\bibinfo{volume}{447}}, \bibinfo{pages}{2603--2630} (\bibinfo{year}{2015}).

\bibitem{Bovy2015}
\bibinfo{author}{{Bovy}, J.}
\newblock \bibinfo{title}{{galpy: A python Library for Galactic Dynamics}}.
\newblock \emph{\bibinfo{journal}{Astrophys. J. Suppl. Ser.}} \textbf{\bibinfo{volume}{216}}, \bibinfo{pages}{29} (\bibinfo{year}{2015}).

\bibitem{Burkert1995}
\bibinfo{author}{{Burkert}, A.}
\newblock \bibinfo{title}{{The Structure of Dark Matter Halos in Dwarf Galaxies}}.
\newblock \emph{\bibinfo{journal}{Astrophys. J. Lett.}} \textbf{\bibinfo{volume}{447}}, \bibinfo{pages}{L25--L28} (\bibinfo{year}{1995}).

\bibitem{Rubin1978}
\bibinfo{author}{{Rubin}, V.~C.}, \bibinfo{author}{{Ford}, J., W.~K.} \& \bibinfo{author}{{Thonnard}, N.}
\newblock \bibinfo{title}{{Extended rotation curves of high-luminosity spiral galaxies. IV. Systematic dynamical properties, Sa -> Sc.}}
\newblock \emph{\bibinfo{journal}{Astrophys. J. Lett.}} \textbf{\bibinfo{volume}{225}}, \bibinfo{pages}{L107--L111} (\bibinfo{year}{1978}).

\bibitem{Noordermeer2007}
\bibinfo{author}{{Noordermeer}, E.}, \bibinfo{author}{{van der Hulst}, J.~M.}, \bibinfo{author}{{Sancisi}, R.}, \bibinfo{author}{{Swaters}, R.~S.} \& \bibinfo{author}{{van Albada}, T.~S.}
\newblock \bibinfo{title}{{The mass distribution in early-type disc galaxies: declining rotation curves and correlations with optical properties}}.
\newblock \emph{\bibinfo{journal}{Mon. Not. R. Astron.}} \textbf{\bibinfo{volume}{376}}, \bibinfo{pages}{1513--1546} (\bibinfo{year}{2007}).

\bibitem{deBlok2008}
\bibinfo{author}{{de Blok}, W.~J.~G.} \emph{et~al.}
\newblock \bibinfo{title}{{High-Resolution Rotation Curves and Galaxy Mass Models from THINGS}}.
\newblock \emph{\bibinfo{journal}{Astron. J.}} \textbf{\bibinfo{volume}{136}}, \bibinfo{pages}{2648--2719} (\bibinfo{year}{2008}).

\bibitem{NestorShachar2023}
\bibinfo{author}{{Nestor Shachar}, A.} \emph{et~al.}
\newblock \bibinfo{title}{{RC100: Rotation Curves of 100 Massive Star-forming Galaxies at z = 0.6-2.5 Reveal Little Dark Matter on Galactic Scales}}.
\newblock \emph{\bibinfo{journal}{Astrophys. J.}} \textbf{\bibinfo{volume}{944}}, \bibinfo{pages}{78} (\bibinfo{year}{2023}).

\bibitem{Weiner2006}
\bibinfo{author}{{Weiner}, B.~J.} \emph{et~al.}
\newblock \bibinfo{title}{{A Survey of Galaxy Kinematics to z\raisebox{-0.5ex}\textasciitilde1 in the TKRS/GOODS-N Field. I. Rotation and Dispersion Properties}}.
\newblock \emph{\bibinfo{journal}{Astrophys. J.}} \textbf{\bibinfo{volume}{653}}, \bibinfo{pages}{1027--1048} (\bibinfo{year}{2006}).

\bibitem{Johnson2021}
\bibinfo{author}{{Johnson}, B.~D.}, \bibinfo{author}{{Leja}, J.}, \bibinfo{author}{{Conroy}, C.} \& \bibinfo{author}{{Speagle}, J.~S.}
\newblock \bibinfo{title}{{Stellar Population Inference with Prospector}}.
\newblock \emph{\bibinfo{journal}{Astrophys. J. Suppl. Ser.}} \textbf{\bibinfo{volume}{254}}, \bibinfo{pages}{22} (\bibinfo{year}{2021}).

\bibitem{Wang2024}
\bibinfo{author}{{Wang}, B.} \emph{et~al.}
\newblock \bibinfo{title}{{The UNCOVER Survey: A First-look HST+JWST Catalog of Galaxy Redshifts and Stellar Population Properties Spanning 0.2 {\ensuremath{\lesssim}} z {\ensuremath{\lesssim}} 15}}.
\newblock \emph{\bibinfo{journal}{Astrophys. J. Suppl. Ser.}} \textbf{\bibinfo{volume}{270}}, \bibinfo{pages}{12} (\bibinfo{year}{2024}).

\bibitem{Chabrier2003}
\bibinfo{author}{{Chabrier}, G.}
\newblock \bibinfo{title}{{Galactic Stellar and Substellar Initial Mass Function}}.
\newblock \emph{\bibinfo{journal}{Publ. Astron. Soc. Pac.}} \textbf{\bibinfo{volume}{115}}, \bibinfo{pages}{763--795} (\bibinfo{year}{2003}).

\bibitem{Charlot2000}
\bibinfo{author}{{Charlot}, S.} \& \bibinfo{author}{{Fall}, S.~M.}
\newblock \bibinfo{title}{{A Simple Model for the Absorption of Starlight by Dust in Galaxies}}.
\newblock \emph{\bibinfo{journal}{Astrophys. J.}} \textbf{\bibinfo{volume}{539}}, \bibinfo{pages}{718--731} (\bibinfo{year}{2000}).

\bibitem{Pacifici2023}
\bibinfo{author}{{Pacifici}, C.} \emph{et~al.}
\newblock \bibinfo{title}{{The Art of Measuring Physical Parameters in Galaxies: A Critical Assessment of Spectral Energy Distribution Fitting Techniques}}.
\newblock \emph{\bibinfo{journal}{Astrophys. J.}} \textbf{\bibinfo{volume}{944}}, \bibinfo{pages}{141} (\bibinfo{year}{2023}).

\bibitem{Burgarella2005}
\bibinfo{author}{{Burgarella}, D.}, \bibinfo{author}{{Buat}, V.} \& \bibinfo{author}{{Iglesias-P{\'a}ramo}, J.}
\newblock \bibinfo{title}{{Star formation and dust attenuation properties in galaxies from a statistical ultraviolet-to-far-infrared analysis}}.
\newblock \emph{\bibinfo{journal}{Mon. Not. R. Astron.}} \textbf{\bibinfo{volume}{360}}, \bibinfo{pages}{1413--1425} (\bibinfo{year}{2005}).

\bibitem{Noll2009}
\bibinfo{author}{{Noll}, S.} \emph{et~al.}
\newblock \bibinfo{title}{{Analysis of galaxy spectral energy distributions from far-UV to far-IR with CIGALE: studying a SINGS test sample}}.
\newblock \emph{\bibinfo{journal}{Astron. Astrophys.}} \textbf{\bibinfo{volume}{507}}, \bibinfo{pages}{1793--1813} (\bibinfo{year}{2009}).

\bibitem{Boquien2019}
\bibinfo{author}{{Boquien}, M.} \emph{et~al.}
\newblock \bibinfo{title}{{CIGALE: a python Code Investigating GALaxy Emission}}.
\newblock \emph{\bibinfo{journal}{Astron. Astrophys.}} \textbf{\bibinfo{volume}{622}}, \bibinfo{pages}{A103} (\bibinfo{year}{2019}).

\bibitem{Yang2022}
\bibinfo{author}{{Yang}, G.} \emph{et~al.}
\newblock \bibinfo{title}{{Fitting AGN/Galaxy X-Ray-to-radio SEDs with CIGALE and Improvement of the Code}}.
\newblock \emph{\bibinfo{journal}{Astrophys. J.}} \textbf{\bibinfo{volume}{927}}, \bibinfo{pages}{192} (\bibinfo{year}{2022}).

\bibitem{Gnerucci2011}
\bibinfo{author}{{Gnerucci}, A.} \emph{et~al.}
\newblock \bibinfo{title}{{Dynamical properties of AMAZE and LSD galaxies from gas kinematics and the Tully-Fisher relation at z \raisebox{-0.5ex}\textasciitilde 3}}.
\newblock \emph{\bibinfo{journal}{Astron. Astrophys.}} \textbf{\bibinfo{volume}{528}}, \bibinfo{pages}{A88} (\bibinfo{year}{2011}).

\bibitem{Livermore2015}
\bibinfo{author}{{Livermore}, R.~C.} \emph{et~al.}
\newblock \bibinfo{title}{{Resolved spectroscopy of gravitationally lensed galaxies: global dynamics and star-forming clumps on {\ensuremath{\sim}}100 pc scales at $1 < z < 4$}}.
\newblock \emph{\bibinfo{journal}{Mon. Not. R. Astron.}} \textbf{\bibinfo{volume}{450}}, \bibinfo{pages}{1812--1835} (\bibinfo{year}{2015}).

\bibitem{Turner2017}
\bibinfo{author}{{Turner}, O.~J.} \emph{et~al.}
\newblock \bibinfo{title}{{The KMOS Deep Survey(KDS): I. Dynamical measurements of typical star-forming galaxies at z = 3.5}}.
\newblock \emph{\bibinfo{journal}{Mon. Not. R. Astron.}} \textbf{\bibinfo{volume}{471}}, \bibinfo{pages}{1280--1320} (\bibinfo{year}{2017}).

\bibitem{Gillman2019}
\bibinfo{author}{{Gillman}, S.} \emph{et~al.}
\newblock \bibinfo{title}{{The dynamics and distribution of angular momentum in HiZELS star-forming galaxies at z = 0.8-3.3}}.
\newblock \emph{\bibinfo{journal}{Mon. Not. R. Astron.}} \textbf{\bibinfo{volume}{486}}, \bibinfo{pages}{175--194} (\bibinfo{year}{2019}).

\bibitem{Swinbank2015}
\bibinfo{author}{{Swinbank}, A.~M.} \emph{et~al.}
\newblock \bibinfo{title}{{ALMA Resolves the Properties of Star-forming Regions in a Dense Gas Disk at z {\ensuremath{\sim}} 3}}.
\newblock \emph{\bibinfo{journal}{Astrophys. J. Lett.}} \textbf{\bibinfo{volume}{806}}, \bibinfo{pages}{L17} (\bibinfo{year}{2015}).

\bibitem{Wu2023}
\bibinfo{author}{{Wu}, Y.} \emph{et~al.}
\newblock \bibinfo{title}{{The Identification of a Dusty Multiarm Spiral Galaxy at z = 3.06 with JWST and ALMA}}.
\newblock \emph{\bibinfo{journal}{Astrophys. J. Lett.}} \textbf{\bibinfo{volume}{942}}, \bibinfo{pages}{L1} (\bibinfo{year}{2023}).

\bibitem{Simons2017}
\bibinfo{author}{{Simons}, R.~C.} \emph{et~al.}
\newblock \bibinfo{title}{{z {\ensuremath{\sim}} 2: An Epoch of Disk Assembly}}.
\newblock \emph{\bibinfo{journal}{Astrophys. J.}} \textbf{\bibinfo{volume}{843}}, \bibinfo{pages}{46} (\bibinfo{year}{2017}).

\bibitem{Wisnioski2015}
\bibinfo{author}{{Wisnioski}, E.} \emph{et~al.}
\newblock \bibinfo{title}{{The KMOS$^{3D}$ Survey: Design, First Results, and the Evolution of Galaxy Kinematics from $0.7 <= z <= 2.7$}}.
\newblock \emph{\bibinfo{journal}{Astrophys. J.}} \textbf{\bibinfo{volume}{799}}, \bibinfo{pages}{209} (\bibinfo{year}{2015}).

\bibitem{Kroupa2001}
\bibinfo{author}{{Kroupa}, P.}
\newblock \bibinfo{title}{{On the variation of the initial mass function}}.
\newblock \emph{\bibinfo{journal}{Mon. Not. R. Astron.}} \textbf{\bibinfo{volume}{322}}, \bibinfo{pages}{231--246} (\bibinfo{year}{2001}).

\bibitem{Maiolino2008}
\bibinfo{author}{{Maiolino}, R.} \emph{et~al.}
\newblock \bibinfo{title}{{AMAZE. I. The evolution of the mass-metallicity relation at $z > 3$}}.
\newblock \emph{\bibinfo{journal}{Astron. Astrophys.}} \textbf{\bibinfo{volume}{488}}, \bibinfo{pages}{463--479} (\bibinfo{year}{2008}).

\bibitem{Salpeter1955}
\bibinfo{author}{{Salpeter}, E.~E.}
\newblock \bibinfo{title}{{The Luminosity Function and Stellar Evolution.}}
\newblock \emph{\bibinfo{journal}{Astrophys. J.}} \textbf{\bibinfo{volume}{121}}, \bibinfo{pages}{161} (\bibinfo{year}{1955}).

\bibitem{Courteau2014}
\bibinfo{author}{{Courteau}, S.} \emph{et~al.}
\newblock \bibinfo{title}{{Galaxy masses}}.
\newblock \emph{\bibinfo{journal}{Reviews of Modern Physics}} \textbf{\bibinfo{volume}{86}}, \bibinfo{pages}{47--119} (\bibinfo{year}{2014}).

\bibitem{Santini2015}
\bibinfo{author}{{Santini}, P.} \emph{et~al.}
\newblock \bibinfo{title}{{Stellar Masses from the CANDELS Survey: The GOODS-South and UDS Fields}}.
\newblock \emph{\bibinfo{journal}{Astrophys. J.}} \textbf{\bibinfo{volume}{801}}, \bibinfo{pages}{97} (\bibinfo{year}{2015}).

\bibitem{Carniani2013}
\bibinfo{author}{{Carniani}, S.} \emph{et~al.}
\newblock \bibinfo{title}{{Strongly star-forming rotating disks in a complex merging system at z = 4.7 as revealed by ALMA}}.
\newblock \emph{\bibinfo{journal}{Astron. Astrophys.}} \textbf{\bibinfo{volume}{559}}, \bibinfo{pages}{A29} (\bibinfo{year}{2013}).

\bibitem{Jones2017}
\bibinfo{author}{{Jones}, G.~C.} \emph{et~al.}
\newblock \bibinfo{title}{{Dynamical Characterization of Galaxies at z {\ensuremath{\sim}} 4-6 via Tilted Ring Fitting to ALMA [C II] Observations}}.
\newblock \emph{\bibinfo{journal}{Astrophys. J.}} \textbf{\bibinfo{volume}{850}}, \bibinfo{pages}{180} (\bibinfo{year}{2017}).

\bibitem{Tadaki2018}
\bibinfo{author}{{Tadaki}, K.} \emph{et~al.}
\newblock \bibinfo{title}{{The gravitationally unstable gas disk of a starburst galaxy 12 billion years ago}}.
\newblock \emph{\bibinfo{journal}{Nature}} \textbf{\bibinfo{volume}{560}}, \bibinfo{pages}{613--616} (\bibinfo{year}{2018}).

\bibitem{Neeleman2019}
\bibinfo{author}{{Neeleman}, M.}, \bibinfo{author}{{Kanekar}, N.}, \bibinfo{author}{{Prochaska}, J.~X.}, \bibinfo{author}{{Rafelski}, M.~A.} \& \bibinfo{author}{{Carilli}, C.~L.}
\newblock \bibinfo{title}{{[C II] 158 {\ensuremath{\mu}}m Emission from z {\ensuremath{\sim}} 4 H I Absorption-selected Galaxies}}.
\newblock \emph{\bibinfo{journal}{Astrophys. J. Lett.}} \textbf{\bibinfo{volume}{870}}, \bibinfo{pages}{L19} (\bibinfo{year}{2019}).

\bibitem{Rizzo2021}
\bibinfo{author}{{Rizzo}, F.}, \bibinfo{author}{{Vegetti}, S.}, \bibinfo{author}{{Fraternali}, F.}, \bibinfo{author}{{Stacey}, H.~R.} \& \bibinfo{author}{{Powell}, D.}
\newblock \bibinfo{title}{{Dynamical properties of z=4.5 dusty star-forming galaxies and their connection with local early-type galaxies}}.
\newblock \emph{\bibinfo{journal}{Mon. Not. R. Astron.}} \textbf{\bibinfo{volume}{507}}, \bibinfo{pages}{3952--3984} (\bibinfo{year}{2021}).

\bibitem{Rizzo2023}
\bibinfo{author}{{Rizzo}, F.} \emph{et~al.}
\newblock \bibinfo{title}{{The ALMA-ALPAKA survey. I. High-resolution CO and [CI] kinematics of star-forming galaxies at z = 0.5-3.5}}.
\newblock \emph{\bibinfo{journal}{Astron. Astrophys.}} \textbf{\bibinfo{volume}{679}}, \bibinfo{pages}{A129} (\bibinfo{year}{2023}).

\bibitem{Fraternali2021}
\bibinfo{author}{{Fraternali}, F.} \emph{et~al.}
\newblock \bibinfo{title}{{Fast rotating and low-turbulence discs at z = 4.5: Dynamical evidence of their evolution into local early-type galaxies}}.
\newblock \emph{\bibinfo{journal}{Astron. Astrophys.}} \textbf{\bibinfo{volume}{647}}, \bibinfo{pages}{A194} (\bibinfo{year}{2021}).

\bibitem{Fujimoto2021}
\bibinfo{author}{{Fujimoto}, S.} \emph{et~al.}
\newblock \bibinfo{title}{{ALMA Lensing Cluster Survey: Bright [C II] 158 {\ensuremath{\mu}}m Lines from a Multiply Imaged Sub-L$^{{\ensuremath{\star}}}$ Galaxy at z = 6.0719}}.
\newblock \emph{\bibinfo{journal}{Astrophys. J.}} \textbf{\bibinfo{volume}{911}}, \bibinfo{pages}{99} (\bibinfo{year}{2021}).

\bibitem{Lelli2021}
\bibinfo{author}{{Lelli}, F.} \emph{et~al.}
\newblock \bibinfo{title}{{A massive stellar bulge in a regularly rotating galaxy 1.2 billion years after the Big Bang}}.
\newblock \emph{\bibinfo{journal}{Science}} \textbf{\bibinfo{volume}{371}}, \bibinfo{pages}{713--716} (\bibinfo{year}{2021}).

\bibitem{Tsukui2021}
\bibinfo{author}{{Tsukui}, T.} \& \bibinfo{author}{{Iguchi}, S.}
\newblock \bibinfo{title}{{Spiral morphology in an intensely star-forming disk galaxy more than 12 billion years ago}}.
\newblock \emph{\bibinfo{journal}{Science}} \textbf{\bibinfo{volume}{372}}, \bibinfo{pages}{1201--1205} (\bibinfo{year}{2021}).

\bibitem{Herrera-Camus2022}
\bibinfo{author}{{Herrera-Camus}, R.} \emph{et~al.}
\newblock \bibinfo{title}{{Kiloparsec view of a typical star-forming galaxy when the Universe was {\ensuremath{\sim}}1 Gyr old. II. Regular rotating disk and evidence for baryon dominance on galactic scales}}.
\newblock \emph{\bibinfo{journal}{Astron. Astrophys.}} \textbf{\bibinfo{volume}{665}}, \bibinfo{pages}{L8} (\bibinfo{year}{2022}).

\bibitem{Nelson2023}
\bibinfo{author}{{Nelson}, E.~J.} \emph{et~al.}
\newblock \bibinfo{title}{{JWST Reveals a Population of Ultrared, Flattened Galaxies at 2 {\ensuremath{\lesssim}} z {\ensuremath{\lesssim}} 6 Previously Missed by HST}}.
\newblock \emph{\bibinfo{journal}{Astrophys. J. Lett.}} \textbf{\bibinfo{volume}{948}}, \bibinfo{pages}{L18} (\bibinfo{year}{2023}).

\bibitem{Parlanti2023}
\bibinfo{author}{{Parlanti}, E.} \emph{et~al.}
\newblock \bibinfo{title}{{ALMA hints at the presence of turbulent disk galaxies at $z > 5$}}.
\newblock \emph{\bibinfo{journal}{Astron. Astrophys.}} \textbf{\bibinfo{volume}{673}}, \bibinfo{pages}{A153} (\bibinfo{year}{2023}).

\bibitem{Roman-Oliveira2023}
\bibinfo{author}{{Roman-Oliveira}, F.}, \bibinfo{author}{{Fraternali}, F.} \& \bibinfo{author}{{Rizzo}, F.}
\newblock \bibinfo{title}{{Regular rotation and low turbulence in a diverse sample of z 4.5 galaxies observed with ALMA}}.
\newblock \emph{\bibinfo{journal}{Mon. Not. R. Astron.}} \textbf{\bibinfo{volume}{521}}, \bibinfo{pages}{1045--1065} (\bibinfo{year}{2023}).

\bibitem{vanderwel2014}
\bibinfo{author}{{van der Wel}, A.} \emph{et~al.}
\newblock \bibinfo{title}{{Geometry of Star-forming Galaxies from SDSS, 3D-HST, and CANDELS}}.
\newblock \emph{\bibinfo{journal}{Astrophys. J. Lett.}} \textbf{\bibinfo{volume}{792}}, \bibinfo{pages}{L6} (\bibinfo{year}{2014}).

\bibitem{Guo2015}
\bibinfo{author}{{Guo}, Y.} \emph{et~al.}
\newblock \bibinfo{title}{{Clumpy Galaxies in CANDELS. I. The Definition of UV Clumps and the Fraction of Clumpy Galaxies at $0.5 < z < 3$}}.
\newblock \emph{\bibinfo{journal}{Astrophys. J.}} \textbf{\bibinfo{volume}{800}}, \bibinfo{pages}{39} (\bibinfo{year}{2015}).

\bibitem{Vega-Ferrero2024}
\bibinfo{author}{{Vega-Ferrero}, J.} \emph{et~al.}
\newblock \bibinfo{title}{{On the Nature of Disks at High Redshift Seen by JWST/CEERS with Contrastive Learning and Cosmological Simulations}}.
\newblock \emph{\bibinfo{journal}{Astrophys. J.}} \textbf{\bibinfo{volume}{961}}, \bibinfo{pages}{51} (\bibinfo{year}{2024}).

\bibitem{Ormerod2024}
\bibinfo{author}{{Ormerod}, K.} \emph{et~al.}
\newblock \bibinfo{title}{{EPOCHS VI: the size and shape evolution of galaxies since z 8 with JWST Observations}}.
\newblock \emph{\bibinfo{journal}{Mon. Not. R. Astron.}} \textbf{\bibinfo{volume}{527}}, \bibinfo{pages}{6110--6125} (\bibinfo{year}{2024}).

\bibitem{Barro2019}
\bibinfo{author}{{Barro}, G.} \emph{et~al.}
\newblock \bibinfo{title}{{The CANDELS/SHARDS Multiwavelength Catalog in GOODS-N: Photometry, Photometric Redshifts, Stellar Masses, Emission-line Fluxes, and Star Formation Rates}}.
\newblock \emph{\bibinfo{journal}{Astrophys. J. Suppl. Ser.}} \textbf{\bibinfo{volume}{243}}, \bibinfo{pages}{22} (\bibinfo{year}{2019}).

\bibitem{Speagle2014}
\bibinfo{author}{{Speagle}, J.~S.}, \bibinfo{author}{{Steinhardt}, C.~L.}, \bibinfo{author}{{Capak}, P.~L.} \& \bibinfo{author}{{Silverman}, J.~D.}
\newblock \bibinfo{title}{{A Highly Consistent Framework for the Evolution of the Star-Forming ``Main Sequence'' from z \raisebox{-0.5ex}\textasciitilde 0-6}}.
\newblock \emph{\bibinfo{journal}{Astrophys. J. Suppl. Ser.}} \textbf{\bibinfo{volume}{214}}, \bibinfo{pages}{15} (\bibinfo{year}{2014}).

\bibitem{Popesso2023}
\bibinfo{author}{{Popesso}, P.} \emph{et~al.}
\newblock \bibinfo{title}{{The main sequence of star-forming galaxies across cosmic times}}.
\newblock \emph{\bibinfo{journal}{Mon. Not. R. Astron.}} \textbf{\bibinfo{volume}{519}}, \bibinfo{pages}{1526--1544} (\bibinfo{year}{2023}).

\end{thebibliography}

\clearpage

\end{document}